The influence of body anisotropy on wake characteristics and enstrophy production for prolate ellipsoids at $Re_D$ = 10,000

Sartaj Tanweer[1†], Mukesh Sharma[1†], Aditya R. Nayak[1, 2], Edwin Malkiel[1], Michael Twardowski[1], Siddhartha Verma[1, 2*]

[1] Harbor Branch Oceanographic Institute, Florida Atlantic University, Fort Pierce, FL 34946, USA.

[2] Department of Ocean and Mechanical Engineering, Florida Atlantic University, Boca Raton, FL 33431, USA.

[†]These authors contributed equally to this work.

*Corresponding author. E-mail: vermas@fau.edu

## Abstract

The flow around prolate ellipsoids is investigated using Large Eddy Simulation (LES) at a Reynolds number of $Re_D$ = 10,000. Five different aspect ratios are considered, with $AR = H/D$ varying from 5:1 to 1:1, where $D$ and $H$ represent the minor- and major- axes, respectively. The major axes of the ellipsoids are set perpendicular to the freestream, and the influence of body anisotropy on boundary layer separation, shear layer behaviour, enstrophy production, and local flow topology is examined. Higher body anisotropy leads to early separation of the boundary layer in the equatorial plane, resulting in a wider wake and a monotonic increase in pressure drag and total drag. Positive enstrophy production reaches a maximum approximately $2.5D$ downstream of the ellipsoids independently of body anisotropy. High body anisotropy leads to sustained negative enstrophy production in the near-wake, specifically near the poles of the 5:1 ellipsoid. Negative production occurs due to the distinct behaviour of streamlines near the high curvature pole, where they undergo strong anisotropic contraction in the cross-stream plane. Interactions between the vorticity vector and the intermediate eigenvector of the strain rate tensor are shown to be the primary source of enstrophy production close to the pole, and the intermediate eigenvalue exhibits negative values in this region. The negative production region is shown to be dominated by the unstable focus / compressing (UF/C) topology, which is consistent with findings from other studies that report negative enstrophy production in turbulent flows.

## 1. Introduction

Boundary layer separation has a significant influence on the characteristics of flow around a body. Depending on the Reynolds number and the shape of the body, the separated shear layers break down via instabilities to give rise to small-scale structures and a highly unsteady wake. A better understanding of the physical mechanisms that play a role in boundary layer detachment and the subsequent evolution of the wake can be valuable when trying to delay, mitigate, or modify the impacts of separation. For this purpose, basic forms like spheres and ellipsoids have long been used as bluff body prototypes for investigating fundamental mechanisms that may apply to more complex geometries of practical importance. Flows around spheres have been studied extensively over a wide range of Reynolds numbers ($Re$) (Achenbach, 1972, 1974; Jang & Lee, 2008; Kim & Pearlstein, 1990; Taneda, 1978; Tomboulides & Orszag, 2000). While many investigations have addressed wake organization, shear layer behaviour, and vortex dynamics for spheres, comparatively fewer studies have systematically examined the influence of shape anisotropy on these features, particularly for separation-dominated flows around bluff bodies. Flows around ellipsoids, specifically spheroids, have received attention predominantly for prolate ellipsoids which are formed by rotating an ellipse about the major axis. The majority of these studies consider scenarios where the major axis is oriented at small to moderate angles of attack with respect to the freestream (Chesnakas & Simpson, 1997; Constantinescu et al., 2002). This setup provides a simplified means of studying complex three-dimensional flow separation over streamlined bodies. A few studies have also considered prolate ellipsoids with the major axis oriented perpendicular to the freestream, a configuration which more closely resembles flow over a bluff body (El Khoury et al., 2010, 2012) and where boundary layer detachment and the separated shear layers play a crucial role in determining overall wake characteristics. A few studies that consider body anisotropy also focus on oblate spheroids formed by rotating an ellipse about the minor axis, due to their geometric similarity to rising bubbles (albeit rigid) (Dandy & Leal, 1986; Magnaudet & Eames, 2000) and to sedimenting particles (Cheng et al., 2024).

An early numerical study by Masliyah & Epstein (1970) used finite difference methods to investigate low Reynolds number flows ($Re$ < 100) for prolate and oblate ellipsoids of various aspect ratios. Magnaudet & Mougin (2007) used direct numerical simulations (DNS) up to $Re$ = 3,000 to determine an empirical criterion for predicting wake stability based on theoretical estimates of the maximum vorticity generated on the surface of the spheroid. Alassar & Badar (1999) conducted Navier-Stokes simulations of oblate ellipsoids of various aspect ratios in the Stokes regime as well as at relatively low Reynolds numbers ($Re$ < 100) to study the time evolution from an impulsive start (Alassar & Badr, 1999). Yang & Prosperetti (2007) performed linear stability analysis of oblate ellipsoids at $Re$ = 150 to 660 for various aspect ratios, to better understand the zigzag path instability of rising bubbles. Li & Zhou (2022) varied the aspect ratio of ellipsoids from 0.5 < $AR$ < 2 (oblate for $AR$ < 1 and prolate for $AR$ > 1) at $Re$ = 300 to observe that the shedding frequency decreased rapidly with increasing $AR$. They also observed an increase in pressure drag and total drag with increasing $AR$. Sanjeevi & Padding (2017) used lattice Boltzmann simulations of oblate and prolate ellipsoids over a range of Reynolds numbers, and varied the angle of attack to show that the drag coefficient at any arbitrary inclination for a given $Re$ could be determined analytically using the drag coefficient values at 0° and 90°. Remarkably, this observation remains valid outside the Stokes regime as demonstrated first by Ouchene et al. (2016) for $Re \leq 240$, and confirmed in the study by Sanjeevi & Padding (2017) primarily for prolate spheroids up to $Re$ = 2000.

At very high Reynolds numbers, a majority of studies focus on slender prolate ellipsoids, often with $AR$ = 6:1 and inclined at various angles of attack to the freestream (Ahn & Simpson, 1992; Vatsa et al., 1989). These studies are motivated by the similarity of the flow to separation over practical slender bodies such as an aircraft fuselage. Tsai & Whitney (1999) used Reynolds Averaged Navier-Stokes (RANS) simulations at $Re = 4.2 \times 10^6$ for angles of attack ranging from 10° to 30°, and found good agreement of surface pressure coefficient and skin friction lines with experimental results from Chesnakas & Simpson (1997) and from Wetzel et al. (1998). Chesnakas & Simpson (1997) used a Laser Doppler Velocimetry (LDV) probe placed inside an inclined prolate spheroid to measure three-component velocity in the inner region of the boundary layer ($y^+ = 7$), to identify separation and re-attachment locations at various cross sections along the body. In addition to static configurations, the 6:1 prolate ellipsoid has also been investigated in dynamic motion both experimentally (Hoang et al., 1994; Taylor et al., 1995; Wetzel & Simpson, 1998) and numerically (Rhee & Hino, 2002; Taylor et al., 1995).

Apart from the extremely high Reynolds number cases discussed above, some studies have examined flows around ellipsoids at relatively low to moderate values of *Re*, where the boundary layer is expected to stay laminar before separation. An early study by Han & Patel (1979) considered the challenges of identifying three-dimensional flow separation using dye lines and tufts on prolate spheroids at various angles of attack. El Khoury et al. (2010) used simulations of a 6:1 spheroid at $Re = 10,000$ and angle of attack 90° to demonstrate that spanwise cross sections of the near-wake resembled the cross section of the ellipsoid. However, the major axis of the wake's cross section became aligned with the minor axis of the spheroid at measurement locations far downstream. Jiang et al. (2016) used DNS of a 6:1 prolate ellipsoid inclined at 45° to describe various stages of wake development for a moderate $Re = 3,000$. They found a distinct coherent vortex tube in the near-wake and observed self-similar variation of velocity deficit in the far-wake. Fröhlich et al. (2020) conducted a series of simulations of flow around prolate spheroids at $1 \leq Re \leq 100$ to develop correlations for lift, drag, and torque using a wide range of aspect ratios and angles of attack.

While the studies discussed above have considered flows around oblate and prolate ellipsoids over a broad range of Reynolds numbers, angles of attack, and aspect ratios, a systematic investigation of the influence of body anisotropy on the vortex dynamics of the flow and its relation to local flow topology has not been carried out at moderate to high Reynolds numbers. More specifically, the influence of body anisotropy has been investigated by a few studies that focus primarily on relatively low Reynolds numbers ($Re \leq 300$). Here, we use LES of crossflow over prolate ellipsoids of various aspect ratios to better understand the influence of a body's shape anisotropy on boundary layer separation, behaviour of the separated shear layers, flow topology, and enstrophy production in the near- and far-wake. The Reynolds number used in the present work is $Re_D = UD/\nu = 10,000$, where $D$ represents the minor axis of the ellipsoid and is kept constant, $U$ is the free-stream velocity, and $\nu$ is the kinematic viscosity of the fluid. This value of *Re* corresponds to the subcritical regime for a sphere in a freestream, i.e., when the boundary layer remains laminar before detachment. The aspect ratio of the prolate ellipsoids is varied from $1 \leq AR = H/D \leq 5$, where $H$ and $D$ represent the major- and minor-axes, respectively. Details of the methodology used are presented in Section 2, and results are discussed in Section 3 followed by conclusion in Section 4.

## 2. Methods
### 2.1 Governing equations

For the Large Eddy Simulations used in the present work, the filtered Navier-Stokes equations have been solved to obtain the spatially filtered velocity field:

$$\frac{\partial \bar{u}_i}{\partial x_i} = 0 \qquad \text{Eq. (1)}$$

$$\frac{\partial \bar{u}_i}{\partial t} + \frac{\partial \bar{u}_i \bar{u}_j}{\partial x_j} = -\frac{\partial \bar{p}}{\partial x_i} + \nu \frac{\partial^2 \bar{u}_i}{\partial x_j \partial x_j} - \frac{\partial \tau_{ij}}{\partial x_j} \qquad \text{Eq. (2)}$$

Here, $\bar{u}_i$ and $\bar{p}$ are the resolved velocity and pressure fields, respectively. The filtering operation on the nonlinear advection term gives rise to the unclosed term $\frac{\partial \tau_{ij}}{\partial x_j}$, where $\tau_{ij} = \overline{u_i u_j} - \bar{u}_i \bar{u}_j$ is referred to as the subgrid-scale (SGS) stress tensor. The SGS stress is modelled as:

$$\tau_{ij} - \frac{\tau_{kk}}{3} \delta_{ij} = -\nu_t \left( \frac{\partial \bar{u}_i}{\partial x_j} + \frac{\partial \bar{u}_j}{\partial x_i} \right) \qquad \text{Eq. (3)}$$

Here, $\nu_t$ is the subgrid-scale turbulent viscosity which is unknown and needs to be modelled. In this study, $\nu_t$ is modelled using the dynamic $k$-equation model (Kim & Menon, 1995), where a transport equation for the SGS kinetic energy ($k_{sgs} = \tau_{ii}/2$) is solved and $\nu_t$ is obtained as:

$$\nu_t = (C_k \Delta)\sqrt{k_{sgs}} \qquad \text{Eq. (4)}$$

Here, $\Delta$ denotes the filter width and $C_k$ is a model constant that is calculated dynamically using a top-hat filter. The filter width is taken as $\Delta V^{\frac{1}{3}}$, where $\Delta V$ is the local cell volume. The overbar notation is omitted for the remainder of the paper, and unless otherwise specified, the resolved velocity and pressure fields are denoted as $u_i$ and $p$.

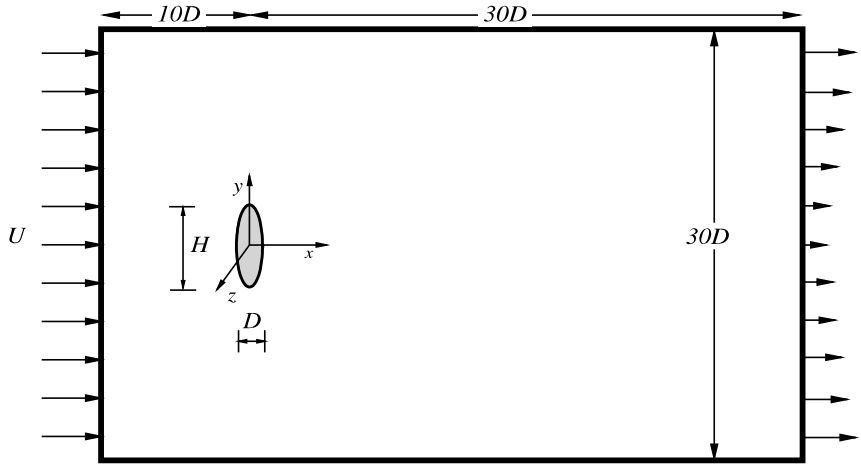

Figure 1: Schematic of crossflow over a prolate ellipsoid. The dimensions of the computational domain are not to scale.

## 2.2 Numerical methods

The governing equations discussed in Section 2.1 were solved using a finite volume based open-source library ('OpenFOAM v11') (Greenshields, 2023; Jasak, 2009).The unsteady terms were discretized using a second-order implicit scheme. The convective and diffusive terms were discretized using a second-order upwind scheme and a second-order central difference scheme, respectively. Linear interpolation was used for interpolating the cell-center values to the face-centers. The pressure-velocity coupling was handled using the Pressure-Implicit with Splitting of Operators (PISO) fractional step algorithm (Issa, 1986).

A schematic of the computational domain is shown in Figure 1. Free-stream velocity was imposed at the inlet as a Dirichlet boundary condition. A zero pressure gradient was imposed at the inlet (Neumann boundary condition) and zero gauge pressure was set at the outlet. An inlet-outlet boundary condition was imposed at the outlet for the velocity as well as for the subgrid-scale kinetic energy (Neumann boundary condition with no reverse flow allowed), and a no-slip boundary condition was imposed at the ellipsoid. Slip boundary conditions were imposed at the lateral boundaries (left, right, top and bottom) which are placed sufficiently far from the ellipsoid and the wake. The time step was selected such that the maximum value of the Courant number ($Co = \Delta t \times \frac{1}{2\Delta V}\Sigma_{faces}|F_i|$) was less than 0.8 to maintain numerical stability. Here, $\Delta t$ denotes the time step, $\Delta V$ is the local cell volume, and $F_i$ represents the volumetric flux at cell faces. In addition to recording instantaneous snapshots for several flow variables at regular intervals, time averaging was started once the flow was fully developed. The total integration time for averaging was longer than $400 U/D$.

To ensure that the meshing and numerical strategies used here simulate the flow accurately, the results obtained for one of the ellipsoids have been assessed in various ways in Appendix A. For maintaining accuracy near solid surfaces, the first cell size in nondimensional wall units (i.e., $y^+ = (u_\tau \Delta y)/\nu$, where $u_\tau$ and $\Delta y$ denote the frictional velocity and the first cell height, respectively) should be kept reasonably small. In direct numerical simulations $y^+$ is typically kept smaller than 1 unit. For one of the ellipsoids used in the present work, the cell heights near the surface (Appendix Figure 3) had a mean value of $y^+ = 1.12$ and a maximum value of $y^+ = 2.78$. Additionally, 80% of turbulent kinetic energy (TKE) should be captured by a well-resolved LES (Pope, 2004). The fraction of resolved TKE ($k_{res} = \bar{u}_i\bar{u}_i/2$) to the total TKE ($k_{res} + k_{sgs}$) is determined as follows:

$$\gamma = \frac{k_{res}}{k_{res} + k_{sgs}} \quad \text{Eq. (5)}$$

The distribution of $\gamma$ around one of the ellipsoids used in this study is shown in Appendix Figure 3(c), and it indicates that more than 80% of TKE is resolved in the majority of the wake. The contribution of the subgrid model in various regions of the flow is also examined using the ratio of $\nu_t/\nu$ for the 5:1 ellipsoid in Appendix Figure 4. Durbin & Reif (2010) suggested that the ratio being smaller than order 10 in the majority of the domain is a useful indicator of a well-resolved LES. It is evident from Appendix Figure 4(a) that the subgrid model has virtually no contribution in the near-wall layer, and minimal contribution in the near-wake. The model's contribution increased modestly in the intermediate-wake and to a greater extent in the far-wake where larger cell sizes were used to manage computational cost (Appendix Figure 4(b)). These tests confirm that the present meshing and numerical strategies are able to represent the flow with reasonable accuracy.

## 3. Results and discussion
### 3.1 Validation

The numerical approach adopted here is validated further by simulating flow around a sphere at $Re_D = UD/\nu$ = 10,000 and comparing the results to data available in the literature (Rodriguez et al., 2019). The results are presented below in nondimensional form, where the length of the minor axis ($D$) and the free-stream velocity ($U$) have been used to normalize relevant variables. The time-averaged streamwise velocity along the centerline in the wake region ($<u_1>$), the turbulence intensity in terms of the root mean square value of the streamwise velocity ($u_{rms}$), and the pressure coefficient ($C_p$) distribution along the surface of the sphere are shown in Figure 2. Here, $C_p$ is defined as follows:

$$C_p = \frac{p - p_\infty}{\frac{1}{2}\rho U^2} \qquad \text{Eq. (6)}$$

where, $p_\infty$ is the pressure at the center line at the inlet boundary, and $\rho$ denotes the density of the fluid. Figure 2 indicates good agreement between the present simulations and the reference data. We note that oscillations in $u_{rms}$ profiles for sphere wakes have been observed by several studies in the streamwise range $1 \leq x/D \leq 3$, even at significantly lower Reynolds numbers. For instance, $u_{rms}$ data from Rodriguez et al. (2011) at $Re_D = 3700$, and from Tomboulides & Orszag (2000) at $Re_D = 300$ display similar oscillations. The authors of these studies note that the maximum in $u_{rms}$ occurs near the tail-end of the recirculation region. This agrees well with the current data shown in Figure 2, where the recirculation region ends at $x/D = 2.05$ (obtained using the zero-crossing of the velocity from Figure 2(a)) and the maximum in $u_{rms}$ occurs at $x/D = 2.045$ (Figure 2(b)).

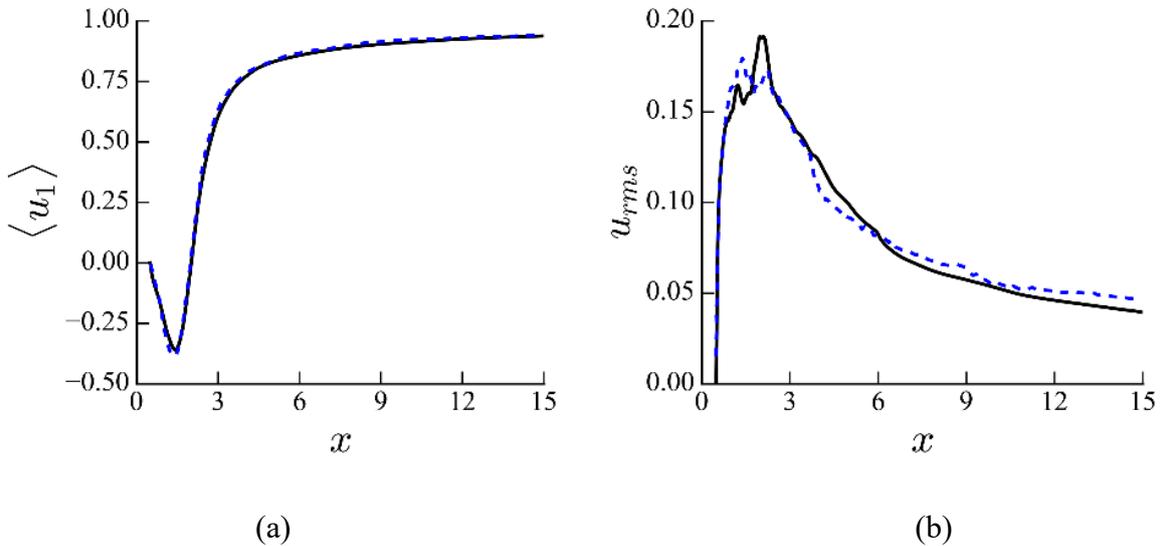

(a)            (b)

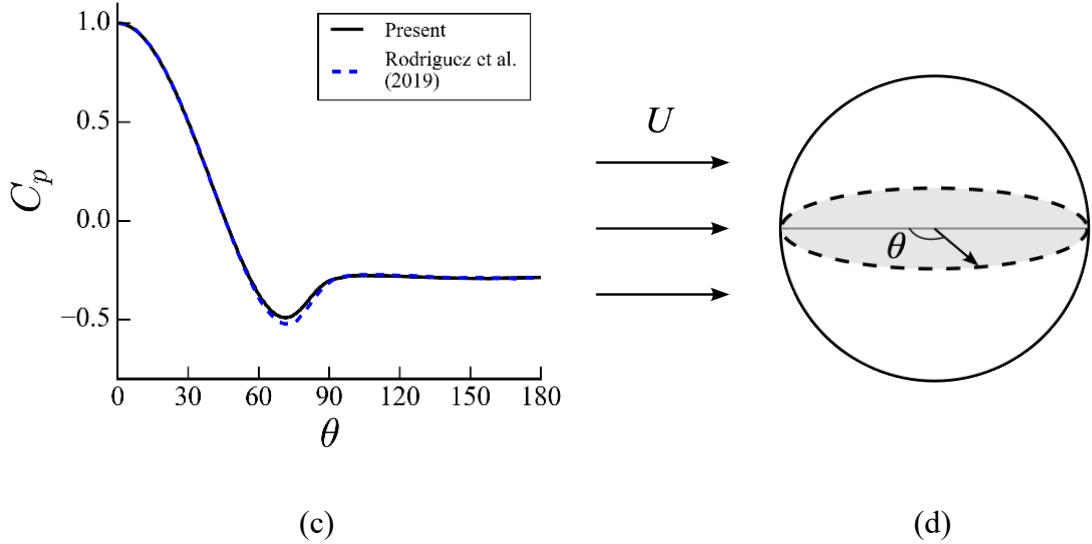

(c)     (d)

Figure 2: Validation for (a) time-averaged streamwise velocity in the wake along the domain center line, (b) root mean square fluctuation of the streamwise velocity, and (c) coefficient of pressure on the sphere surface. In these plots, *x* denotes the downstream distance from the center of the ellipsoid normalized with *D*, and (d) $\theta$ is the azimuthal angle starting from the upstream stagnation point.

### 3.2 Boundary layer separation

Unlike for a sphere, the cross-sections and flow patterns for ellipsoids with non-unity values of *AR* differ in the meridional (*x-y*) and equatorial (*x-z*) planes. To account for these differences introduced by body anisotropy, the flow behavior is examined and compared for various combinations of $AR = H/D$ = 5:1, 5:2, 5:3, 5:4, and in some instances for 1:1 (i.e., a sphere), in the meridional and equatorial planes. These orthogonal planes are shown in Figure 3 for the 5:1 ellipsoid. All the results discussed here are presented in nondimensional form, where *U* and *D* have been used to nondimensionalize the relevant quantities. Various properties within the boundary layer are examined first, along with the formation of the separated shear layers. The enstrophy transport equation is examined in a later section to study the phenomena of enstrophy generation and depletion in the separated flow regions and in the wake. The interactions between the vorticity vector and the strain rate tensor are also examined at a later point to better understand the origin of enstrophy generation and depletion.

Boundary layer separation is typically estimated to occur at the location where surface shear stress becomes zero. While this criterion provides a reasonable estimate for quasi-2D flow separation, such as for crossflow over cylinders with uniform cross section, it may not always provide a complete description of three-dimensional separation (Maskell, 1955; Wang, 1972). Separation in three-dimensions may occur through other mechanisms, such as through vortex roll-up observed on prolate spheroids inclined at moderate angles of attack. In such scenarios, the helicity density has been shown to be a useful metric for identifying separation locations (Chesnakas & Simpson, 1997; Plasseraud & Mahesh, 2025). Notably, for large angles of attack, separation leads to the formation of a large circulation bubble as is typically observed in quasi-2D crossflow over circular cylinders (Wang, 1972). The angle of attack for the ellipsoids in the present work is effectively 90° with respect to the major axis, and a large separation bubble forms in the wake as opposed to the rollup type separation and reattachment observed for moderate angles of attack. Moreover, the boundary layer detaches along a separation line or curve that varies continuously with the surface curvature of the ellipsoids, instead of separating at a nearly constant mean elevation angle as in the case of cylinders.

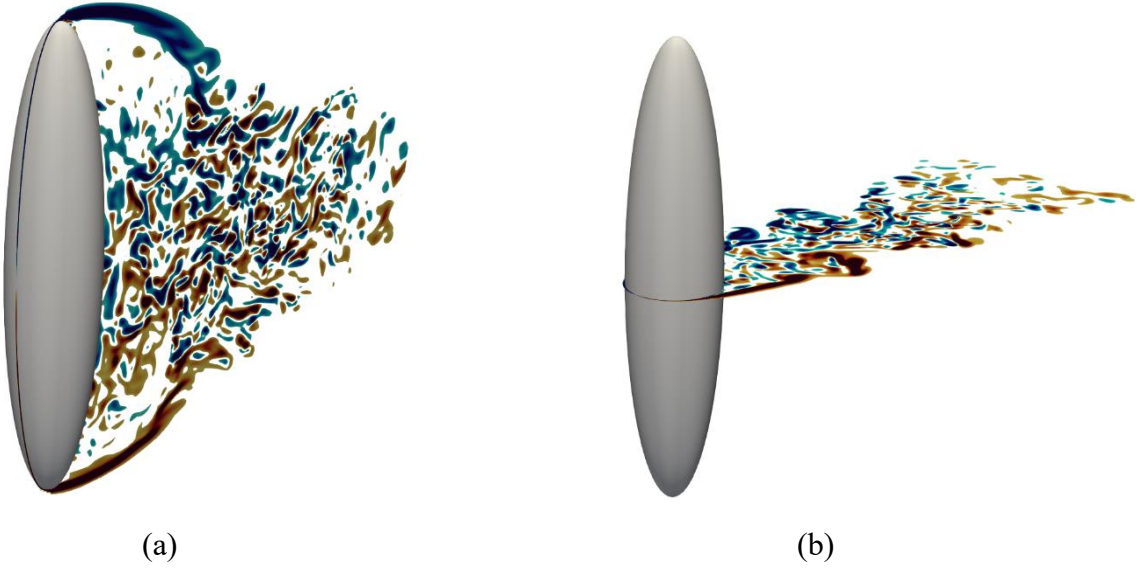

(a)                (b)

Figure 3: View of (a) the meridional (*x-y*) and (b) equatorial (*x-z*) planes passing through the center of the 5:1 ellipsoid. The instantaneous *z*-vorticity and *y*-vorticity are shown in (a) and (b), respectively (red - positive, blue - negative).

Wetzel et al. (1998) carried out a detailed investigation of three-dimensional separation using various experimental techniques, including oil flow visualization, surface pressure measurements, skin-friction measurements, and laser Doppler velocimetry. They demonstrated that even for complex three-dimensional separation scenarios, minima in wall shear stress magnitude correlate well with separation locations. Thus, separation is identified in the present work by locating minima in plots of wall (or surface) shear stress magnitude. For this purpose, the instantaneous value of shear stress was calculated as $\tau_i = T_i - (T_j . n_j)n_i$. Here, $T_i$ denotes the traction vector calculated using the deviatoric part of the stress tensor ($T_i = \mu\left(\frac{\partial u_i}{\partial x_j} + \frac{\partial u_j}{\partial x_i}\right)n_j$), $n_j$ denotes the surface normal unit vector, and the wall shear stress magnitude is given as $\tau_w = |\tau_i|$. The wall shear stress was time-averaged for a duration of $400U/D$, and resulting plots for $\tau_w$ are shown in Figure 4 for both the meridional and equatorial planes for all 5 ellipsoids. The corresponding numerical values for elevation (α) and azimuthal (θ) separation angles, calculated using the minima in the $\tau_w$ curves, are provided in Table 1. We observe that the elevation separation angle decreases with decreasing *AR* ratios, which indicates earlier separation of the boundary layer for increasing body symmetry (or correspondingly, for lower body anisotropy). The sphere undergoes earliest separation at elevation angle $\alpha \approx 85^0$ in the meridional plane, whereas the most elongated 5:1 ellipsoid undergoes later separation at $\alpha \approx 89.6^0$.

To better understand the observed separation behaviour, the pressure coefficient $C_p = (p - p_\infty)/(1/2\,\rho U^2)$ in the meridional (x-y) plane along the upper surface of the ellipsoids is shown in Figure 5(a). For smaller AR ratios (5:3, 5:4 and 1:1), the flow experiences a favorable pressure gradient (FPG) along the surface up until $\alpha \approx 75^0$; a gradual decrease in pressure is observed up to this point followed by a gradual increase (adverse pressure gradient - APG). Due to the FPG, the flow in the boundary layer gets accelerated, leading to an initial increase in surface shear stress in the meridional plane (Figure 4(a)). The subsequent APG region (Figure 5(a)) causes the flow to decelerate, and the surface shear stress decreases (Figure 4(a)). Due to this adverse pressure gradient, the boundary layer undergoes separation at some point, and a constant surface pressure is observed downstream of this location (i.e., beyond $\alpha \approx 90^0$ in Figure 5(a)).

For larger AR ratios (i.e., 5:1 and 5:2) the pressure initially decreases along the surface up until $\alpha \approx 80^0$ (Figure 5(a)). Beyond this point the flow undergoes an abrupt rise in pressure. The steep decrease and increase in pressure for $\alpha > 60°$, when compared to more gradual pressure changes observed for smaller AR ratios (5:3, 5:4 and 1:1), are a result of the sharp curvature near the upper and lower poles of the 5:1 and 5:2 ellipsoids. This geometric feature first gives rise to a large FPG and subsequently to a large APG. The APG leads to boundary layer separation at some point, beyond which the pressure undergoes minor fluctuations. These fluctuations are not observed for the smaller AR ratios (5:3, 5:4 and 1:1) discussed previously, for which the post-separation surface pressure asymptotes to constant values. The plots also suggest a correlation between wall shear stress magnitude and surface curvature. Regions of high curvature, e.g., $\alpha > 60°$ for the 5:1 and 5:2 ellipsoids, experience stronger flow acceleration and hence higher wall shear stress magnitude (Figure 4(a)), whereas regions of low curvature experience comparatively lower shear stress due to weaker flow acceleration. Overall, these observations provide an initial indication of how the severity of body anisotropy influences boundary layer behaviour, separation angle, and post-separation characteristics in the meridional plane.

For considering separation characteristics in the equatorial plane, we note that the equatorial cross sections for all 5 ellipsoids are geometrically identical regardless of the AR ratio, since the minor-axis diameter is kept constant. This geometric similarity may explain why the profiles of $\tau_w$ plotted against the azimuthal angle $\theta$ do not show significant variation with AR in Figure 4(b). However, the azimuthal separation angle $\theta$, determined using the locations of minima in Figure 4(b), exhibits a measurable increase with decreasing AR ratio (Table 1). This is contrary to the trend observed in the meridional plane where separation occurs earlier for lower body anisotropy (i.e., for the sphere), but separation in the equatorial plane occurs earlier for higher body anisotropy (i.e., for the most elongated ellipsoid). The 5:1 ellipsoid undergoes separation at $\theta \approx 76.4^0$ in the equatorial plane, whereas the sphere undergoes separation at $\theta \approx 85.2^0$. This observation can be related to $C_p$ reaching its minima earlier for higher AR ratios in Figure 5(b). Unlike the abrupt pressure changes observed in the meridional plane for high AR values, the pressure varies gradually in the equatorial plane for all values of AR (Figure 5(b)). This can be related to the surface curvature remaining constant in the equatorial plane, whereas rapid curvature changes occur in the meridional plane near the poles for the 5:1 and 5:2 ellipsoids.

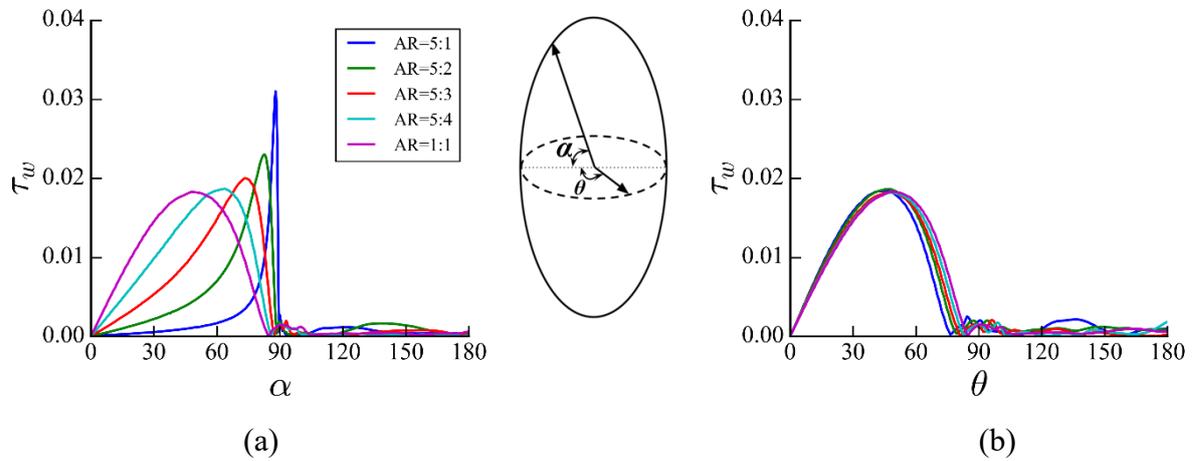

(a)                                      (b)

Figure 4: The magnitude of surface shear stress ($\tau_w$) in (a) the meridional plane (*x-y*) and (b) the equatorial plane (*x-z*). Here $\alpha$ and $\theta$ represent the elevation and azimuthal angles, respectively, and they vary from $0^0$ to $180^0$.

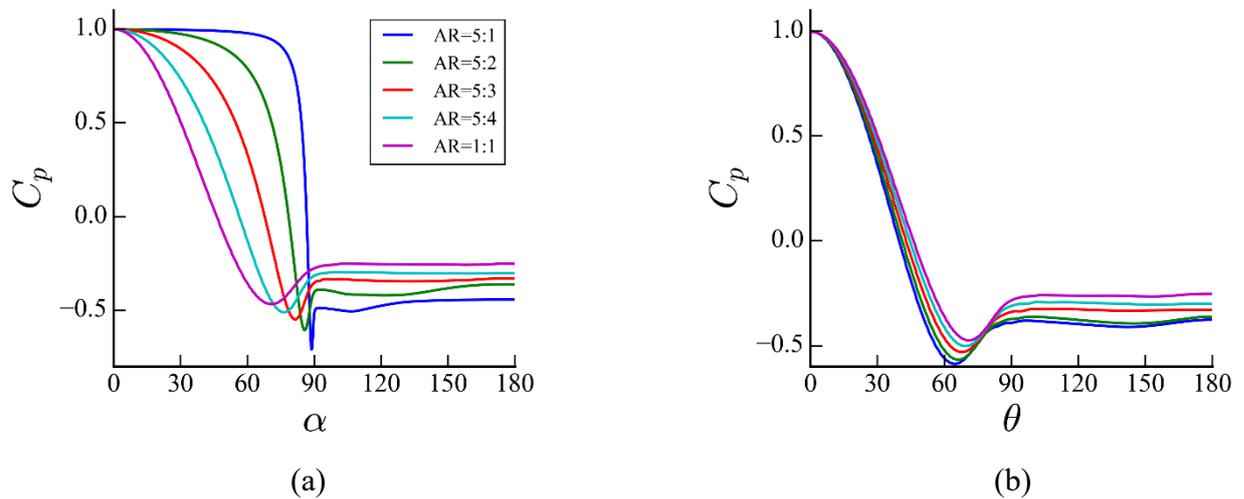

(a)                                      (b)

Figure 5: Surface pressure coefficient in (a) the meridional plane (*x-y*) and (b) the equatorial plane (*x-z*). Here $\alpha$ and $\theta$ correspond to the elevation and azimuthal angles, respectively, as shown in the schematic in Figure 4.

Table 1: Separation angle from the upstream stagnation point (in degrees) in the meridional and equatorial planes, and time-averaged drag coefficients for various $AR$ values.

| Aspect ratio ($AR$) | 5:1 | 5:2 | 5:3 | 5:4 | 1:1 |
|---|---|---|---|---|---|
| Elevation separation angle ($\alpha$) | 89.6 | 88.6 | 87.4 | 86.0 | 85.2 |
| Azimuthal separation angle ($\theta$) | 76.4 | 79.4 | 81.4 | 82.7 | 85.2 |
| Drag coefficient ($C_D$) | 0.701 | 0.610 | 0.519 | 0.472 | 0.410 |
| Pressure drag coefficient ($C_{D,p}$) | 0.677 | 0.585 | 0.492 | 0.430 | 0.376 |
| Viscous drag coefficient ($C_{D,v}$) | 0.024 | 0.025 | 0.027 | 0.042 | 0.034 |

In addition to considering separation angles in the meridional and equatorial planes, flow separation over the entire surface of the ellipsoids can be visualized in three-dimensions (3D) by locating minima in wall shear stress. Figure 4 indicates that the wall shear stress is close to zero at the minima locations. Thus, contours of zero wall shear stress (excluding the forward stagnation point) were used to define separation curves on the ellipsoids' surface. To determine whether these contours are consistent with separation locations, limiting streamlines were calculated for the 5:1 ellipsoid and the corresponding results are shown in Appendix Figure 2. The zero-stress contour aligns closely with the convergence of the limiting streamlines, which demonstrates that the separation curves provide a reasonable indication of boundary layer detachment.

To better understand the influence of body anisotropy on boundary layer detachment, the separation curves for the 5:1 and 5:4 ellipsoids are compared in Figure 6. These curves appear as narrow bands instead of lines since a small range of shear stress values, rather than a single zero value, was used to locate the minima. This approach helps mitigate the effects of noise near the separation location. In addition to the 3D representations shown in Figure 6(a) and Figure 6(b), 2D projections of the points that generate the separation curves are shown in Figure 6(c) and Figure 6(d). These projections were obtained by considering various horizontal cross sections parallel to the equatorial plane and identifying the azimuthal separation angle in each such cross section. To depict the resulting information on the polar plot, the distance of each cross section from the upper pole of the ellipsoid (normalized by $D$) was used to mark radial distance on the polar plot, and the azimuthal angle was used to mark the polar angle. The resulting plots show a continuous increase in azimuthal separation angle for both ellipsoids as we move from the equator to the poles. However, higher body anisotropy corresponds to comparatively earlier separation for horizontal cross sections farther from the pole. We observe more noticeable differences between the two ellipsoids closer to the equator than to the poles, which is somewhat unexpected since the equatorial cross sections are geometrically identical between the 5:1 and 5:4 ellipsoids. The early separation closer to the equator for the 5:1 ellipsoid is an indirect consequence of a larger wake signature in the height dimension (major axis), which in turn leads to a larger overall wake envelope in 3D and consequently a broader wake in the width dimension (minor axis). The broader wake width corresponds to the boundary layer separating earlier as we move closer to the equator.

Quantitative differences in wake-height and -width for the two ellipsoids are examined in Figure 7. Although the shear layers form a 3D surface around the periphery of the ellipsoids, 2D cross sections can offer valuable insights regarding the influence of body anisotropy on the shape and width of the wake. The wake size in the two orthogonal planes is examined using the plane-normal vorticity components along cross-stream line cuts located $1.5D$ downstream of the ellipsoid centers. The maxima/minima in vorticity occur near the core of the shear layers, and the distance between the extrema is smaller for the 5:4 ellipsoid than for the 5:1 ellipsoid. This indicates that the shear layer of the 5:4 ellipsoid diverges away from the wake centerline to a smaller degree, resulting in a narrower wake.

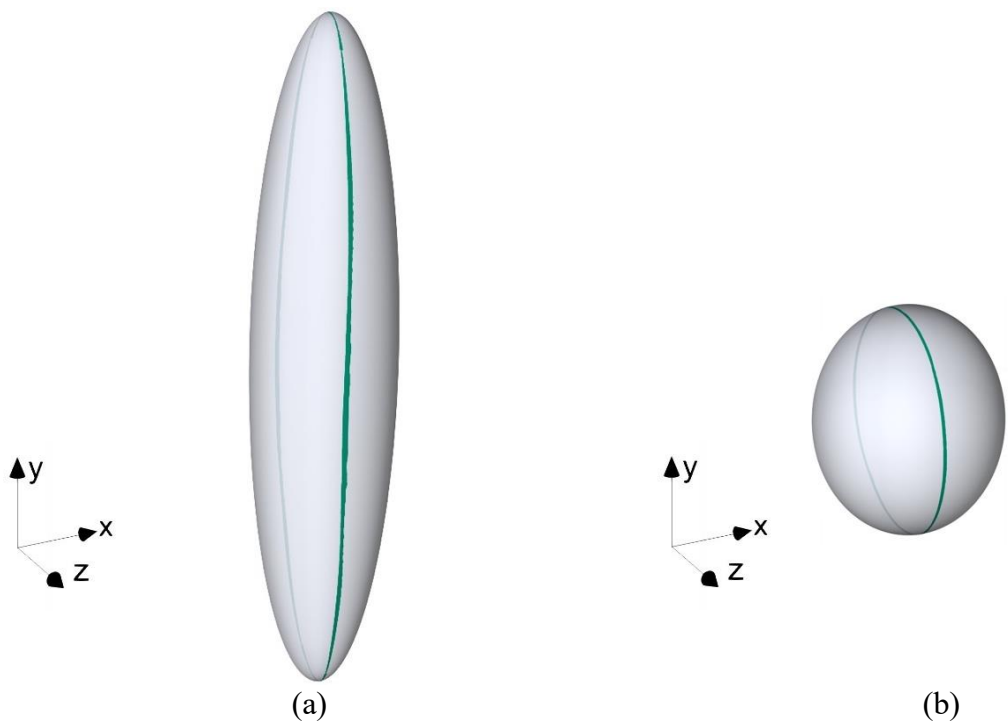

(a)  (b)

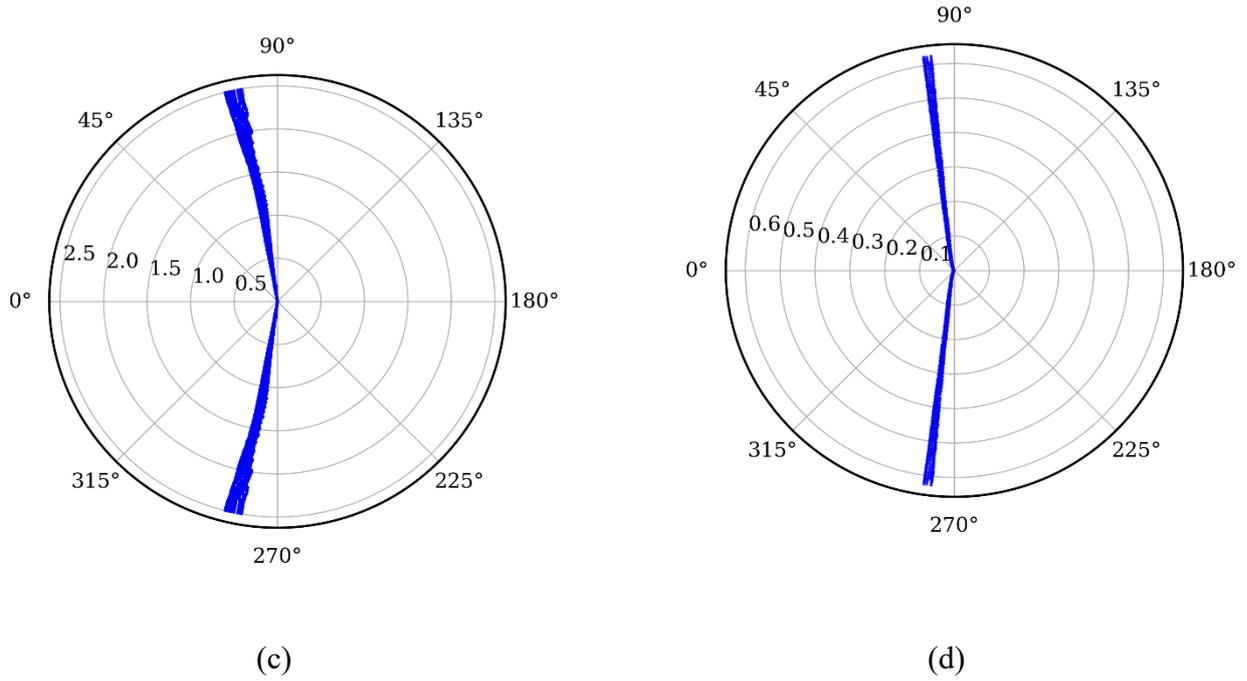

(c)              (d)

Figure 6: (a, b) Separation curves and (c, d) polar distribution of the corresponding points in various cross sections parallel to the equatorial plane, shown for the 5:1 ellipsoid (a,c) and the 5:4 ellipsoid (b,d). The radial distance in the polar plots indicates non-dimensional vertical distance from the major axis poles of the respective ellipsoids. The polar angle is measured with respect to the upstream stagnation point on the ellipsoids.

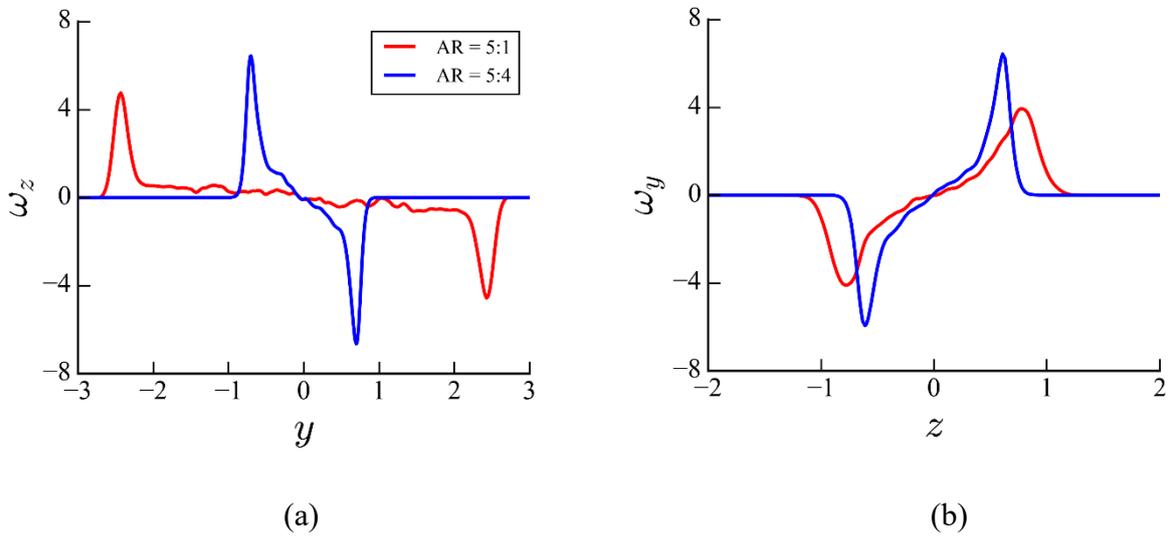

(a)              (b)

Figure 7: (a) Time-averaged nondimensional $z$-vorticity in the meridional plane and (b) $y$-vorticity in the equatorial plane, taken from line cuts located at $x = 1.5D$ downstream of the ellipsoid centers.

### 3.3 Wake velocity and drag coefficient

As the boundary layer separates and the 3D shear layer is formed, a recirculation region develops in the near-wake of the ellipsoids. Velocity profiles taken along line-cuts in the wake provide an indication of the size of these recirculation regions as well as of other general wake characteristics, albeit only in a single dimension along the line cut. The time-averaged streamwise velocity profiles ($< u_1 >$) shown in Figure 8(a) indicate that for all 5 ellipsoids considered, there is an initial flow reversal corresponding to negative velocity values, followed by a recovery to various levels that are lower than the freestream velocity. The size of the recirculation region can be estimated using the location where the velocity first recovers to zero. It is evident that the size increases monotonically with increasing aspect ratio, and that the most elongated 5:1 ellipsoid generates a recirculation bubble that is substantially larger than the rest. The velocity profiles also provide an indication of the intensity of recirculation, with more negative minima corresponding to stronger recirculation bubbles. The trend in this regard suggests an increase in recirculation intensity with increasing aspect ratio.

The velocity defect along the wake centerline, defined as $u_d = (U - u_1)/U$, is shown in Figure 8(b). Two distinct decay rates of the time-averaged $< u_d >$ are observed for all cases except the 5:1 ellipsoid. For the sphere with $AR$=1:1, the decay of $< u_d >$ exhibits a $x^{-2}$ power law in the near-wake region for $1.33D \leq x \leq 3D$ (with $x$ measured from the downstream surface), followed by a slower decay of $x^{-3/4}$ for $x \geq 4D$. A rapid decay in velocity defect indicates faster recovery towards freestream conditions, and it is typically associated with a smaller wake and lower drag. A slower initial decay rate is indicative of slow recovery, a large wake, and higher drag. The decrease in decay rate beyond $x \geq 4D$ for the sphere is a result of the wake velocity having already recovered substantially within $x \leq 3D$, which is followed by a very gradual asymptote towards the freestream value (Figure 8(a)). The 5:1 ellipsoid displays a consistent $x^{-1.13}$ decay rate for the majority of the wake, and the remaining asymmetric ellipsoids exhibit decay rates that lie between this case and that of the sphere. This trend in velocity defect suggests that the drag coefficient is higher for increasing body anisotropy, which is confirmed by the calculated drag values provided in Table 1. The transition from two distinct decay rates to a single decay rate with increasing aspect ratio can be attributed to differences in the wake structure. As discussed previously, the 5:1 ellipsoid gives rise to a larger wake not just in the meridional plane but also in the equatorial plane. This can also be observed in Supplementary Movie 1, which depicts the normal vorticity components in these two orthogonal planes. The influence of flow recirculation persists farther downstream for the 5:1 ellipsoid, resulting in a consistent but weaker decay of the velocity defect. For the 5:4 ellipsoid, which closely resembles the sphere but retains some degree of anisotropy, the influence of the recirculation region is limited in comparison, as can be observed in Supplementary Movie 1. Thus, the streamwise velocity recovers rapidly beyond $x > 1.5D$ (Figure 8(a)) giving rise to a faster initial decay rate, followed by a slow asymptote towards the freestream value.

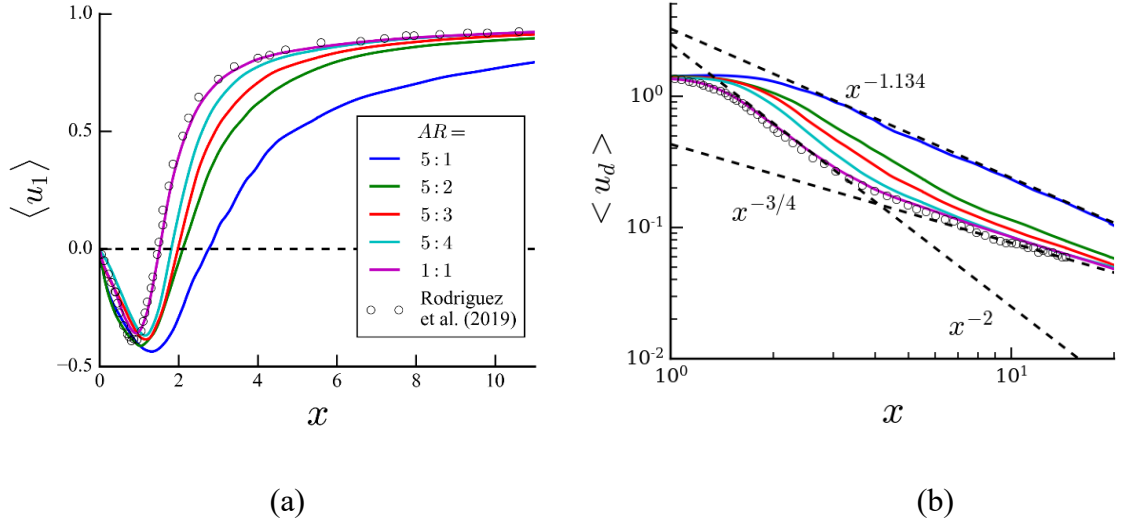

(a)  (b)

Figure 8: (a) Time-averaged streamwise velocity and (b) velocity defect along the centerline in the wake ($y = 0$, $z = 0$). The horizontal axis indicates non-dimensional distance from the rear surface of the ellipsoids. The dot symbols correspond to reference LES data from Rodriguez et al. (2019). The intersection of the dashed line in (a) with the solid lines provides an indication of the size of the recirculation bubble.

Apart from the wake velocity profiles discussed above, the variation in drag coefficient with body anisotropy is also examined. The time-averaged drag coefficient $C_D = F_D/(\frac{1}{2}\rho U^2 A)$ for all 5 ellipsoids is provided in Table 1. $F_D$ represents the net drag force and $A$ represents the projected frontal area ($A = \pi DH/4$). The contributions of viscous (or skin-friction) drag and pressure-induced drag to the total drag coefficient have also been listed in Table 1. We observe that pressure drag shows an increasing trend for higher $AR$ values. This is a result of the gauge pressure (not shown) in the near-wake being considerably more negative for higher values of $AR$ than for lower values. A larger pressure difference between the upstream and downstream surface leads to higher pressure drag. The wake pressure can also be related indirectly to the velocity defect decay rates discussed previously; a slower decay rate signifies a larger wake, and consequently, an extensive region of low wake pressure. Some variations are observed in the skin-friction drag coefficient, and its contribution is an order of magnitude smaller than that of pressure drag at the given Reynolds numbers. The magnitude of the pressure drag being dominant leads to a monotonic increase in total $C_D$ with increasing values of $AR$.

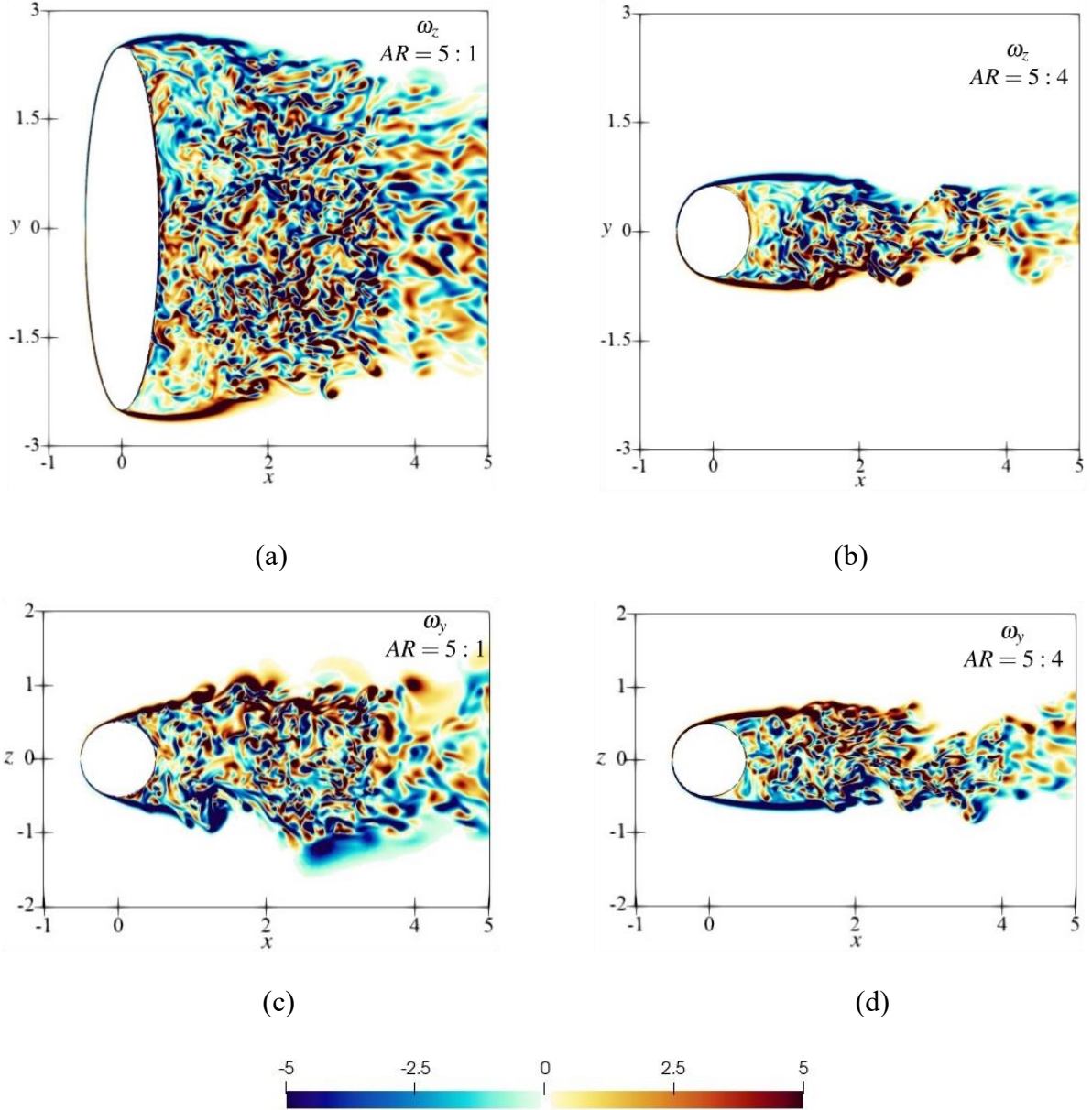

Figure 9: (a, b) Instantaneous $z$-vorticity in the meridional plane, and (c, d) instantaneous $y$-vorticity in the equatorial plane. The panels on the left (a, c) correspond to the 5:1 ellipsoid, and those on the right (b, d) correspond to the 5:4 ellipsoid. The vorticity has been nondimensionalized by dividing with $U/D$, and the axes have been nondimensionalized using $D$. An animation of this figure is provided in Supplementary Movie 1.

**3.4 Shear layer characteristics and flow topology**

The separated shear layers are examined using instantaneous vorticity contours in the meridional and the equatorial planes in Figure 9. The current and subsequent sections focus primarily on the 5:1 and 5:4 ellipsoids since these two shapes represent the highest and lowest non-zero body anisotropy, respectively. For both the ellipsoids, the separated shear layers were observed to remain coherent for a small distance downstream before undergoing roll-up, with subsequent breakdown giving rise to a broad range of spatial and temporal scales in the wake. The roll-up locations were identified using instantaneous vorticity contours, where a roll-up event was defined as the point where the shear layers in the two orthogonal plane cuts transitioned from a relatively flat profile to forming a discrete rotating material element. The center coordinates of these discrete material elements were identified prior to pinch-off, and the values were averaged over a duration of 10 non-dimensional time units to obtain the roll-up locations shown in Figure 10. We observe that the roll-up process in the meridional plane starts at comparable distances downstream of the ellipsoid centers for both *AR* values, specifically between $x/D \approx 0.8$ to 1 on average. Moreover, there is very little variation of the roll-up location in the *y* direction, as is evident from the barely perceptible vertical error bars for the red symbols. In the equatorial plane, roll-up occurs earlier for the 5:1 ellipsoid at around $x/D \approx 0.3$ on average, but the shear layer stays coherent farther downstream for the 5:4 ellipsoid, up to approximately $x/D \approx 0.8$. There is significant variation in both the streamwise (*x*) and cross-stream (*z*) location of roll-up in the equatorial plane, which matches qualitative observations from Supplementary Movie 1.

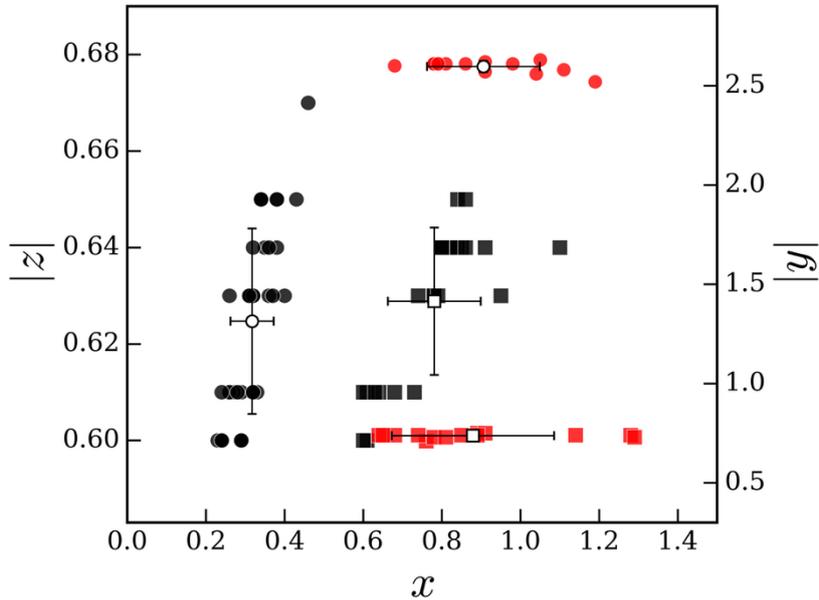

Figure 10: Shear layer roll-up locations for the 5:1 ellipsoid (circles) and the 5:4 ellipsoid (squares), recorded for multiple roll-up events in the equatorial plane (black) and in the meridional plane (red). The values have been non-dimensionalized using *D*, and they indicate distance from the ellipsoids' centers. The cross-stream coordinate locations are plotted along the left vertical axis for the equatorial plane ($|z|$), and along the right vertical axis for the meridional plane ($|y|$). The mean values are plotted using open symbols, and the error bars indicate the standard deviation in both directions.

To examine strong rotational structures that form as the shear layers evolve in space and time, coherent structures in the wakes of the 5:1 and 5:4 ellipsoids are visualized using the *Q*-criterion in Figure 11 (also shown in Supplementary Movie 2). Larger *Q* values are associated with stronger rotational structures, and the strongest structures appear closer to the equator than to the poles for both the ellipsoids. The near-wakes of the ellipsoids are largely devoid of *Q* structures, although additional structures would become visible if contours with smaller *Q* values were used. The absence of $Q$ structures suggests that the strongest tube-like rotational structures are typically absent in the near-wake where the shear layers remain cohesive. However, early breakdown of the shear layer in the equatorial plane for the 5:1 ellipsoid (discussed previously) leads to strong near-wake vortex tubes forming close to the equator in Figure 11(a). The strongest vortex tubes (i.e., contours with larger $Q$ values) form at a certain standoff distance, approximately $2.5D$ downstream of the ellipsoid centers for both the cases shown in Figure 11. This can be related to enstrophy production reaching its maximum around this downstream distance, which will be discussed later in Section 3.5.

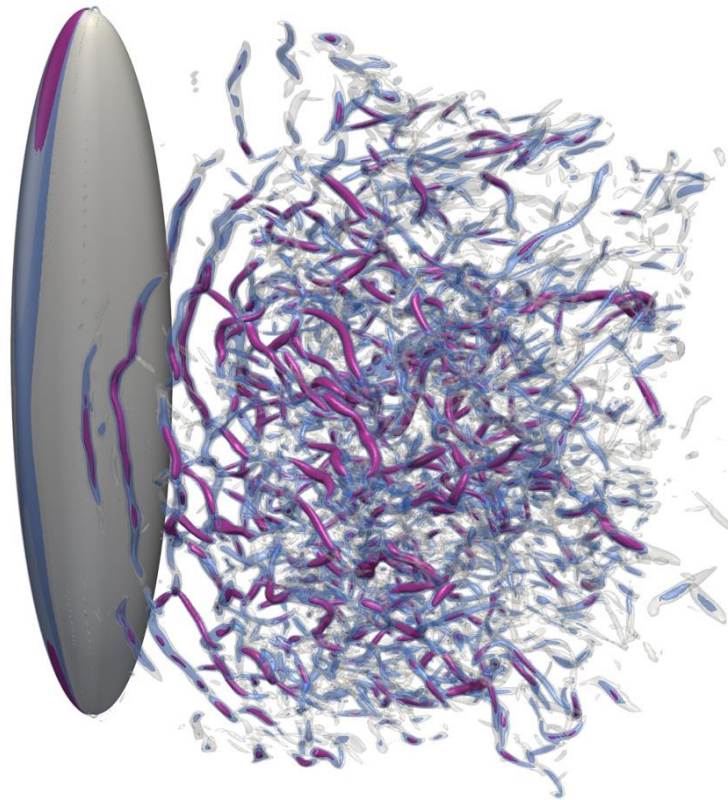

(a)

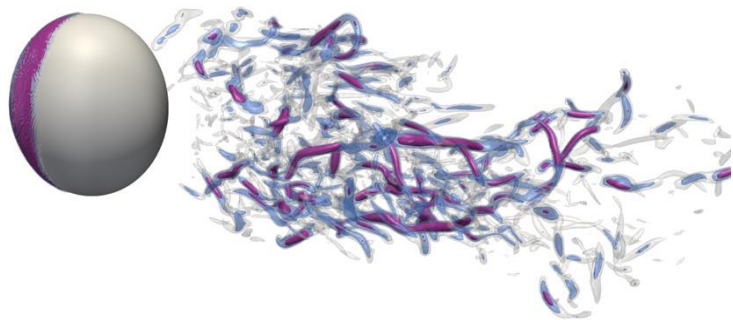

(b)

Figure 11: Isosurfaces of the *Q*-criterion for (a) the 5:1 ellipsoid and (b) the 5:4 ellipsoid. The colours shown correspond to normalized values of $Q/(U/D)^2 = 50$ (gray), 100 (blue), and 200 (red). An animation of this figure is provided in Supplementary Movie 2.

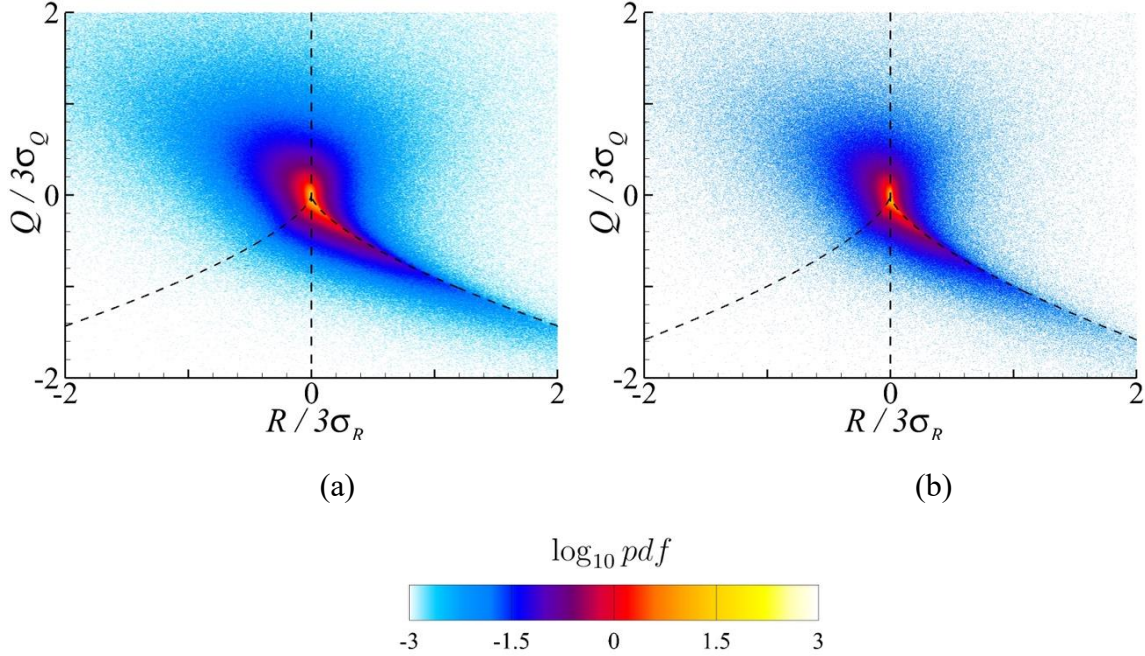

(a)  (b)

$\log_{10} pdf$

Figure 12: Joint probability density function of *Q* and *R* for the (a) 5:1 ellipsoid and the (b) 5:4 ellipsoid. Both *Q* and *R* have been normalized using three times their respective standard deviations ($\sigma_Q$ and $\sigma_R$).

The characteristics of the resolved small-scale wake structures are examined in Figure 12 using joint probability density functions (PDF) of the second and third invariants (*Q* and *R*) of the velocity gradient tensor (Chong et al., 1990; Ooi et al., 1999; Soria et al., 1994). Additional details regarding the relationship of *Q* and *R* to the local flow topology are provided in Appendix B. We observe in Figure 12 that the *Q-R* joint PDFs for the 5:1 and 5:4 ellipsoids bear some similarities, and both display a characteristic 'teardrop' shape which is typically observed in a variety of turbulent flow configurations. The teardrop shape was observed consistently across all 5 ellipsoids examined in this study, with minor influence of body anisotropy on the distribution of points. This indicates that despite notable differences in geometry, the resolved scales in the wakes display characteristics that are typical of universal small-scale behaviour in turbulent flows. Some noticeable differences between the two cases shown in Figure 12 are a broader spread of the distribution for the 5:1 ellipsoid, as well as a higher probability of encountering larger values of *Q* and *R*. The spread can be quantified using the standard deviation values $\sigma_Q$ and $\sigma_R$, which are 1.47 times and 1.53 times larger for the 5:1 ellipsoid than for the 5:4 ellipsoid. The mean value of *Q* is positive for both cases, with $\mu_Q$ being 1.78 times larger for the 5:1 ellipsoid. This suggests that vorticity dominates over strain in the wakes of both ellipsoids, but that the vortex tubes are considerably stronger for the 5:1 ellipsoid. Overall, these observations provide a quantitative indication that higher body anisotropy in cross flow leads to the formation of stronger rotational coherent structures, which matches qualitative observations from Figure 11.

## 3.5 Enstrophy production

The spatial and temporal evolution of wake structures are linked closely to the production and dissipation of vorticity, and thus, to the production of enstrophy. Enstrophy is defined as $\xi = 0.5|\nabla \times \boldsymbol{u}|^2$ and it provides a quantitative measure of vorticity strength. The transport equation for enstrophy is provided in Appendix C, and the main focus of this section is on the enstrophy production term $\omega_i S_{ij} \omega_j$. Using eigendecomposition of the symmetric matrix $S_{ij}$, this term can be decomposed into contributions from the three eigenvalues $s_i$ of the strain-rate tensor, and the alignment between the corresponding eigenvectors and the vorticity vector (Buxton & Ganapathisubramani, 2010):

$$\xi_{prod} = \omega_i S_{ij} \omega_j = \omega^2 s_i (\widehat{\boldsymbol{\omega}} \cdot \boldsymbol{e}_i)^2 \qquad \text{Eq. (7)}$$

Here, $\omega$ represents the magnitude of vorticity, $\widehat{\boldsymbol{\omega}}$ is the unit vector in the direction of vorticity, and $\boldsymbol{e}_i$ are the three eigenvectors of the strain-rate tensor. When $s_i$ is arranged in ascending order, the first eigenvalue $s_1$ is always negative and the third eigenvalue $s_3$ is always positive. The sum of the three eigenvalues must be zero to satisfy the continuity equation. For the enstrophy production term $\omega^2 s_i (\widehat{\boldsymbol{\omega}} \cdot \boldsymbol{e}_i)^2$ in Eq. (7), we observe that only the term $s_i$ can be negative. Hence, the positive eigenvalue $s_3$ always contributes to enstrophy generation whereas the negative eigenvalue $s_1$ always contributes to enstrophy depletion. The intermediate eigenvalue $s_2$ can contribute locally either to enstrophy generation or to depletion, depending on whether its value is positive or negative. A similar eigendecomposition was used by Watanabe et al., (2014) in DNS of a planar turbulent jet to demonstrate that positive enstrophy production becomes less dominant along the upstream-oriented (trailing) edge of the turbulent/non-turbulent interface. This results in a marked reduction in enstrophy generation along the trailing edge, due to an increased prevalence of vortex compression events associated with local negative enstrophy production. Nevertheless, the net enstrophy production along the trailing edge was observed to remain weakly positive.

The expression for enstrophy production in Eq. (7) can be simplified further as $\xi_{prod} = \omega^2(s_1 \cos^2 \phi_1 + s_2 \cos^2 \phi_2 + s_3 \cos^2 \phi_3)$, where $\cos \phi_i = \widehat{\boldsymbol{\omega}} \cdot \boldsymbol{e}_i$ are the direction cosines of the vorticity vector in the eigenframe. It is known that the vorticity vector aligns preferentially with the intermediate eigenvector $\boldsymbol{e}_2$ in homogeneous isotropic turbulence, and that it is orthogonal to the compressive eigenvector $\boldsymbol{e}_1$ and shows no preferential alignment with the extensive eigenvector $\boldsymbol{e}_3$ (Ashurst et al., 1987). Preferential alignment with $\boldsymbol{e}_2$ is observed almost universally across a broad range of turbulent flow configurations. While this leads to the direction cosine $\cos \phi_2$ being the largest among the three, the intermediate eigenvalue $s_2$ is known to be the smallest in magnitude. Thus, it is difficult to attribute the contribution of any one of the three eigenvalues or eigenvectors as being dominant to enstrophy production.

The behavior of the three direction cosines and the corresponding eigenvalues is analyzed later in this section to better understand their individual roles.

Figure 13 shows the normalized time-averaged enstrophy production $P_\xi = <\omega_i S_{ij} \omega_j>/(U/D)^3$ in the meridional and equatorial planes for the four asymmetric ellipsoids used in this study. The most prominent differences appear in the meridional planes due to greatest dissimilarity of the corresponding cross section profiles. For low body anisotropy, i.e., for the 5:3 and 5:4 ellipsoids, we observe shear layers from opposite poles merging into a single region of high enstrophy production approximately $2.5D$ downstream of the ellipsoid centers. With increasing body anisotropy, the high production region begins to separate out towards the poles, and two distinct regions are visible for the 5:1 ellipsoid in the meridional plane. Another notable difference in the meridional planes is the presence of negative production regions for the highly anisotropic 5:1 and 5:2 ellipsoids close to the poles. However, $P_\xi$ remains positive beyond $x/D > 1.5$ for these ellipsoids, indicating that the flow primarily undergoes vortex stretching in the intermediate- and far-wake regardless of body anisotropy. It is known that positive enstrophy production dominates overall in turbulent flows due to the prevalence of vortex stretching, which is related to the forward energy cascade. However, enstrophy production can be negative in localized regions, corresponding to strong vortex compression (Bechlars & Sandberg, 2017; Buxton & Ganapathisubramani, 2010). As discussed in Appendix C, negative production indicates a local reduction in enstrophy and a corresponding suppression of strong vortex tubes. This can be confirmed qualitatively from Figure 11 and from Supplementary Movie 2, where large $Q$-value contours are absent close to the upper and lower poles of the 5:1 ellipsoid.

Unlike in the meridional plane, enstrophy production is primarily positive in the equatorial plane for all four ellipsoids, indicating a prevalence of vortex stretching regardless of the degree of body anisotropy. Similar to the behaviour observed in the meridional plane, the shear layers merge into a single high production region for low body anisotropy in the equatorial plane, but they remain separated into two distinct regions for higher body anisotropy. Overall, there are significant differences visible in the equatorial plane among the 4 ellipsoids, even though all 4 equatorial cross sections are geometrically identical. The magnitude of enstrophy production appears to increase noticeably with increasing body anisotropy, which can be correlated qualitatively to the presence of strong rotational structures (i.e., contours of large $Q$-value) near the equator in Figure 11(a). Higher enstrophy production also appears to correlate well qualitatively with the shear layer being more susceptible to instabilities resulting in roll-up and breakdown, as can be seen in Figure 9(c).

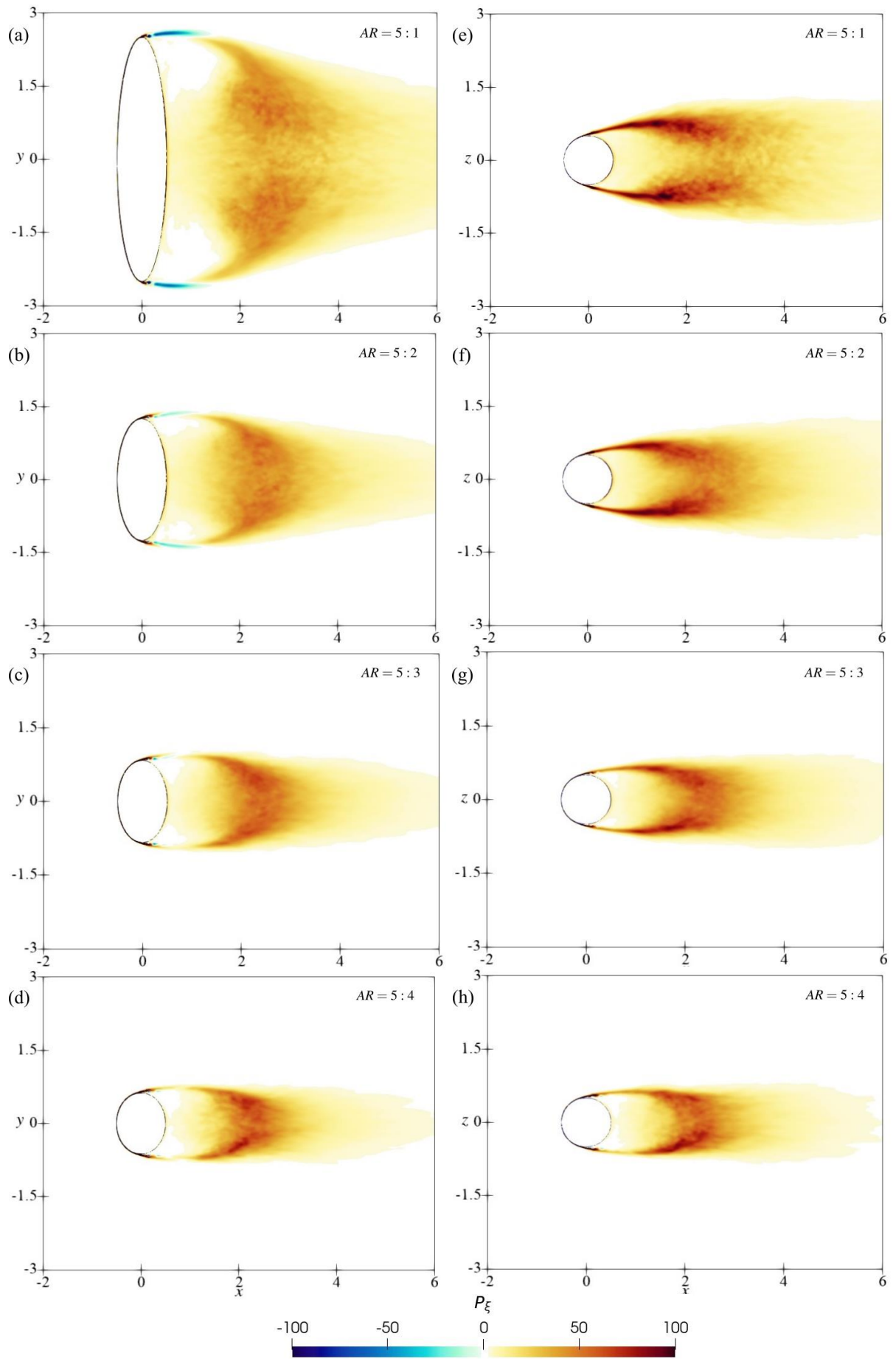

Figure 13: Time-averaged enstrophy production $P_\xi = <\omega_i S_{ij} \omega_j>/(U/D)^3$ in (a-d) the meridional planes, and (e-h) the equatorial planes for the four asymmetric ellipsoids. The values for enstrophy production have been normalized by dividing with $(U/D)^3$ and the axes have been normalized using $D$.

While the 2D plane cuts shown in Figure 13 provide a preliminary indication of the spatial distribution of enstrophy production, examining 3D structures can provide a more comprehensive picture of the origin of negative enstrophy production and of large positive production. Figure 14 shows various contours of the positive time-averaged enstrophy production for the 5:1 and 5:4 ellipsoids. Three distinct contour levels with non-dimensional production values $P_\xi = 10, 100$, and 200 have been visualized in the figure. We observe that for both ellipsoids, regions of highest positive $P_\xi$ (red contours) are generally concentrated in the core of the wake, approximately 2.5$D$ downstream from the ellipsoids' centers. It also appears that the positive $P_\xi$ contours exhibit shell-like structures, with high magnitude $P_\xi$ contours nested within low magnitude $P_\xi$ contours. The strongest $P_\xi$ shells do not connect to the ellipsoids' surface, indicating that positive enstrophy production reaches its maximum in the intermediate wake. As the value of $P_\xi$ decreases, the shells extend towards the ellipsoids' surface, with the lowest $P_\xi = 10$ shells wrapping around the surface, enveloping the boundary layer and the separated shear layer. The most prominent difference between the highly anisotropic 5:1 ellipsoid and the low anisotropy 5:4 ellipsoid is the separation of the $P_\xi = 200$ shell into two distinct lobes closer to the poles for the 5:1 ellipsoid, which was discussed previously using 2D plane cuts in Figure 13. Overall, Figure 14 indicates that positive enstrophy production remains relatively low in the near-wake region regardless of body anisotropy. This is followed by a gradual increase, reaching a maximum around $x$=2.5$D$ for the cases shown here, followed by a gradual decline further downstream.

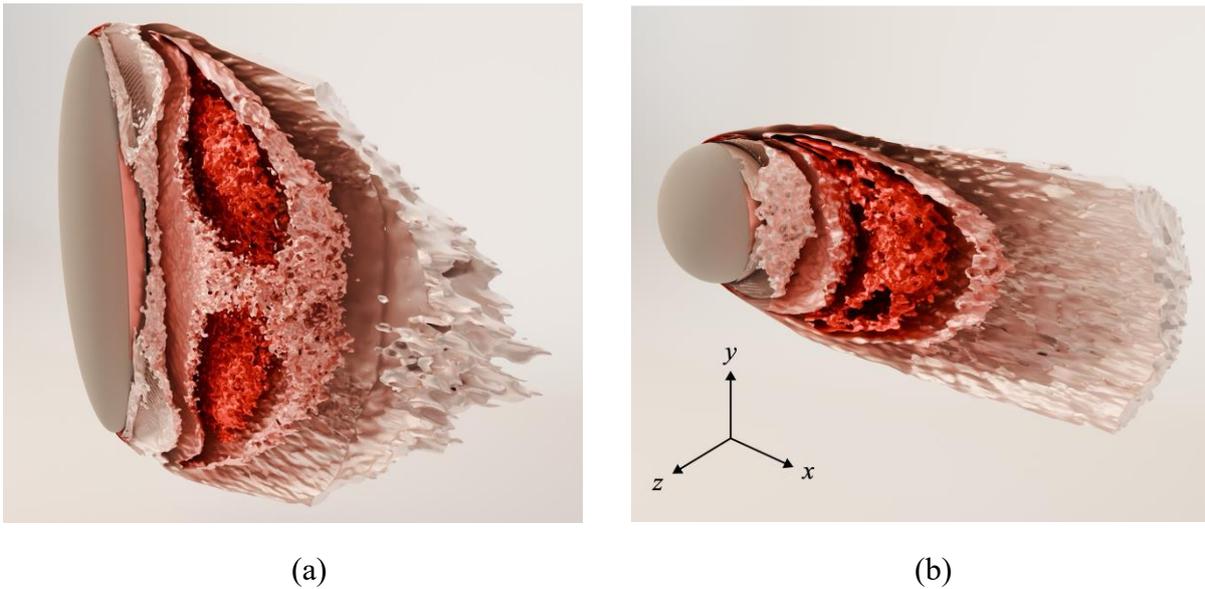

(a)            (b)

Figure 14: (a) Contours showing positive time-averaged enstrophy production in the wake of the 5: 1 ellipsoid. The contours have been cut away along the meridional plane to show the internal three-dimensional structure of positive production, and the contours correspond to non-dimensional values of $P_\xi = 10$ (white), 100 (pink), and 200 (red). (b) Positive enstrophy production for the 5:4 ellipsoid, with the same contour levels as shown in panel (a).

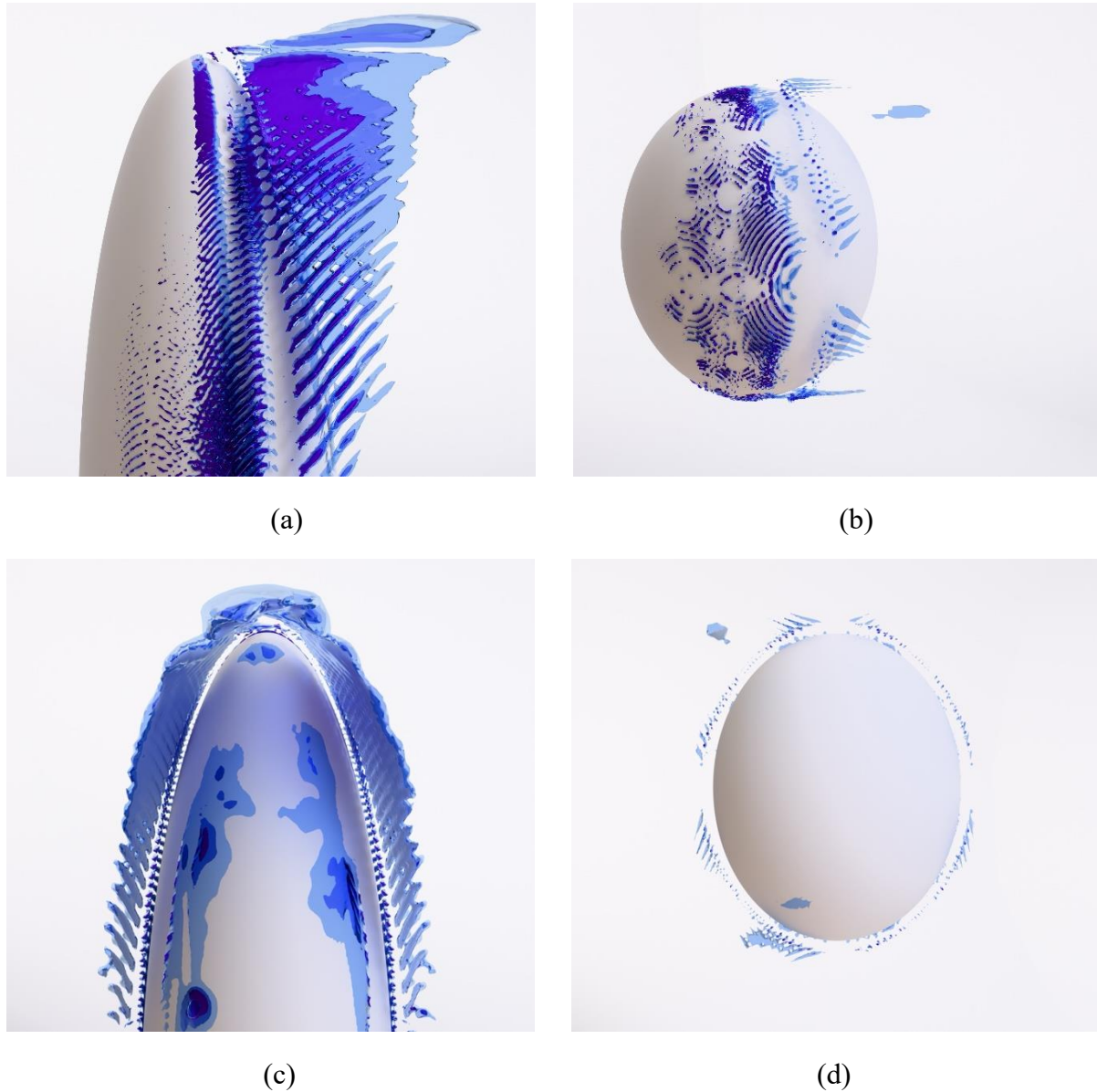

Figure 15: (a) Side view of regions of negative time-averaged enstrophy production for the 5:1 ellipsoid, with contour levels corresponding to $P_\xi = -10$ (light blue), $-50$ (blue), and $-100$ (violet). (b) Negative production for the 5:4 ellipsoid, with the same contour levels as in panel (a). Panels (c) and (d) show downstream views of the images in panels (a) and (d).

The contours of negative time-averaged enstrophy production for the 5:1 and 5:4 ellipsoids are shown in Figure 15. For both the ellipsoids, $P_\xi \leq -10$ is constrained to thin regions around the periphery. For the 5:1 ellipsoid, the negative production region takes the form of a hood-like structure, with a large and prominent canopy forming near the poles that diminishes towards the equator. The finger-like striations observed closer to the equator arise from disproportionate amplification of discretization errors when computing enstrophy production, since $\omega_i S_{ij} \omega_j$ is cubic in velocity gradients. However, such oscillations are not present in the hood-like canopy close to the pole of the ellipsoid. Similarly to the positive production shells shown in Figure 14, the negative production contours with higher $P_\xi$ magnitude are nested within lower magnitude $P_\xi$ contours. However, differently from positive production, negative time-averaged production attains its maximum magnitude in the near-wake close to the ellipsoid surface, and not in the intermediate wake. It is also evident from Figure 15 that negative production regions for the 5:4 ellipsoid are essentially absent. This suggests that high body anisotropy may be more conducive to negative enstrophy production being sustained over extended periods of time in localized spatial regions. Sustained negative production is not a common occurrence in incompressible turbulent flows since vortex stretching is expected to be dominant in general. Thus, the source of negative production for the 5:1 ellipsoid is investigated in further detail.

To determine whether the length scales associated with persistent negative production regions are substantially larger than the LES filter scale, an instantaneous velocity snapshot was filtered a-posteriori using a Gaussian kernel of filter width 16Δ. Here, Δ is the local cell size defined as the cube root of each cell's volume. Instantaneous negative enstrophy production regions were calculated using the filtered velocity field, and the hood-like canopy near the pole of the 5:1 ellipsoid remained clearly visible, indicating that this feature resides on length scales substantially larger the LES filter width. This large-scale nature also suggests that the persistent negative production region is only weakly influenced by the subgrid model, which is consistent with the minimal model contribution observed in the near-wake in Appendix Figure 4.

To better understand the cause of sustained negative production near the poles of the 5:1 ellipsoid, streamlines were generated near the upper pole using an instantaneous velocity snapshot. A spanwise line of tracer particles was released upstream of the ellipsoid, and the resulting streamlines are shown coloured using the normalized instantaneous enstrophy production $\xi_{prod}/(U/D)^3$ in Figure 16(a) and Figure 16(b). A relatively large cohesive region of instantaneous negative production is visible in the near-wake region in Figure 16(b), and it has been highlighted using a green rectangle. As described earlier, enstrophy production can be decomposed into the three contributing terms $\omega^2 s_i \cos^2 \phi_i$, and only the compressive and intermediate eigenvectors $s_1$ and $s_2$ can give rise to negative production since $s_3$ is always positive. Figure 16(c) shows probability density functions of the magnitude of the three direction cosines $\cos \phi_i$ within the highlighted rectangular region. It is evident that the eigenvectors $\boldsymbol{e}_1$ and $\boldsymbol{e}_3$ are both almost perfectly orthogonal to the vorticity vector in this region, whereas the intermediate eigenvector $\boldsymbol{e}_2$ shows near perfect alignment with vorticity. This indicates that enstrophy production in the highlighted region is determined almost entirely by the intermediate eigenvector term $\omega^2 s_2 \cos^2 \phi_2$. Thus, the intermediate eigenvalue is examined in Figure 16(d) by colouring the streamlines with $s_2$. The positive and negative values of $s_2$ show a near perfect correspondence with the sign of enstrophy production (Figure 16(b)) in the highlighted region. This is expected, since $\omega^2 s_2 \cos^2 \phi_2$ is the only term with a meaningful contribution to enstrophy production, as discussed above.

To assess whether the long-term behaviour of the direction cosines $\cos\phi_i$ matches their instantaneous behaviour, time-averaged values of $|<\cos\phi_i>|$ were examined in the negative enstrophy region defined by $P_\xi \leq -10$ in Figure 15. Similarly to the instantaneous case, the magnitudes of time-averaged compressive and extensive direction cosines were found to be significantly smaller than that of the intermediate direction cosine. This indicates that over the long term, vorticity remains nearly perfectly aligned with the intermediate eigenvector within the highlighted region of interest, and it remains orthogonal to both the compressive and extensive eigenvectors. Consequently, the long-term enstrophy production is expected to be governed by the same $\omega^2 s_2 \cos^2\phi_2$ term that dominates instantaneous production as described above.

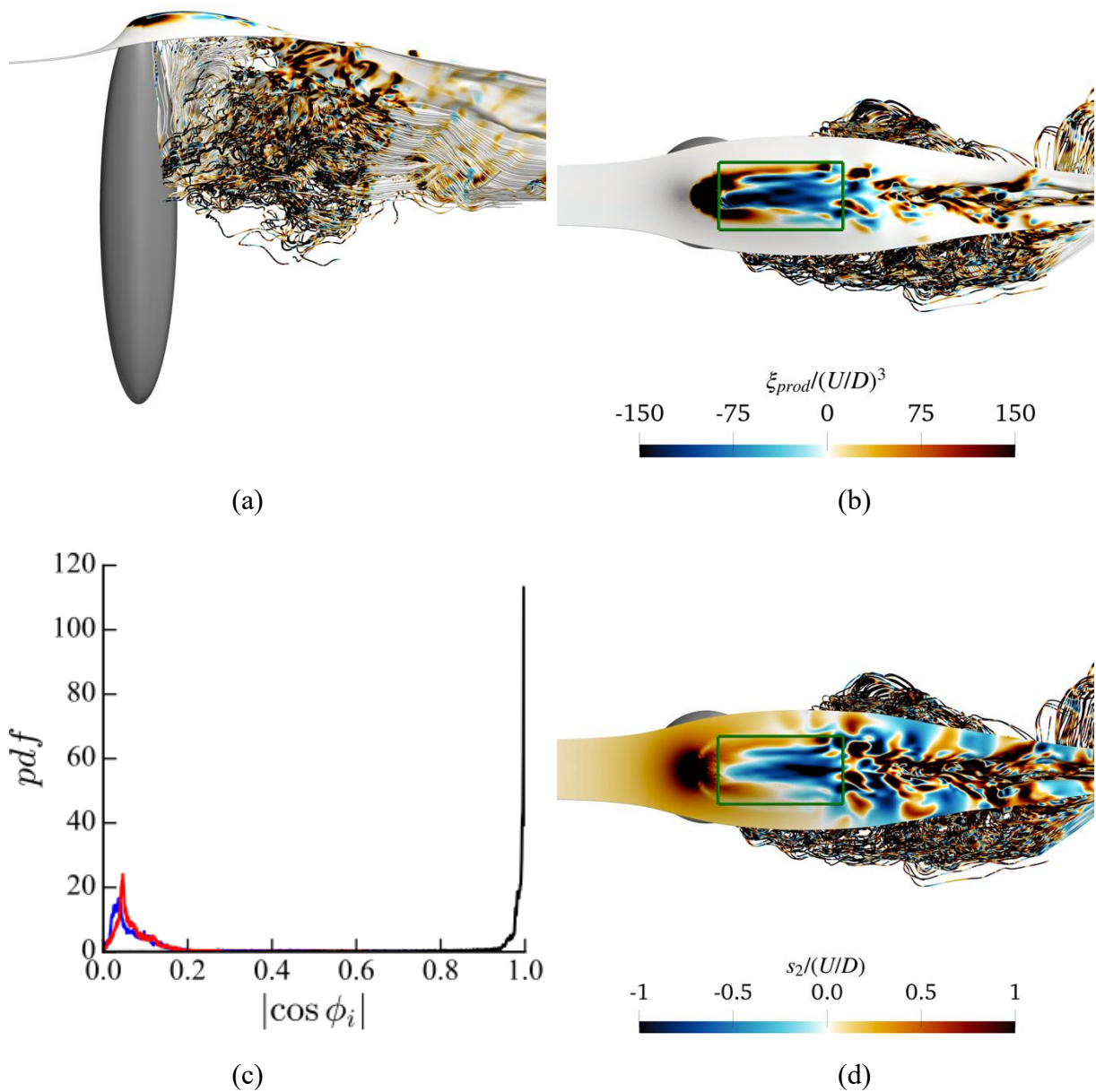

Figure 16: (a) Side view of streamlines generated by tracer particles released upstream of the 5:1 ellipsoid in a single velocity snapshot. The streamlines have been coloured using the instantaneous normalized enstrophy production $\xi_{prod}/(U/D)^3$ (colourbar shown in panel (b)). (b) Top view of the streamlines from panel (a). A green rectangle highlights the region where a large cohesive region of negative production is observed. (c) Probability density functions of the magnitude of the three direction cosines in the highlighted rectangular region, with the line for $|cos\,\phi_1|$ shown in blue, $|cos\,\phi_2|$ shown in black, and $|cos\,\phi_3|$ shown in red. (d) Top view of the streamlines coloured with the intermediate eigenvector $s_2$ normalized using *U/D*.

The eigenvalue $s_2$ being negative in the region of interest (Figure 16(d)) indicates that the flow experiences compression in two orthogonal directions (since $s_1 < 0$ always) and stretching in one orthogonal direction (since $s_3 > 0$ always). In Figure 16, as the horizontal streamlines approach the highly anisotropic upper pole of the 5:1 ellipsoid, they experience severe deflection in the vertical (*y*) direction (Figure 16(a)) in addition to lateral deflection in the spanwise (*z*) direction (Figure 16(b)). Immediately downstream of the ellipsoid, the streamlines are pulled back in sharply from multiple directions into the near-wake, before evolving into convoluted trajectories; the streamlines at the upper pole bend downward in the vertical direction and they contract inward in the spanwise direction. This pronounced cross-stream wake contraction manifests as the flow experiencing two orthogonal compressive eigenvalues, i.e., as $s_1$ and $s_2$ both being negative in the highlighted region of interest. The severity of streamline bending was quantified by calculating the curvature of streamlines passing close to the upper pole as $\boldsymbol{\kappa} = \boldsymbol{t} \cdot \nabla \boldsymbol{t}$, where $\boldsymbol{t} = \boldsymbol{u}/|\boldsymbol{u}|$ is the unit velocity vector. The $\kappa_y$ component of curvature was observed to be strongly negative within the highlighted rectangular region shown in Figure 16, which provides quantitative confirmation that the streamlines bend downward sharply, resulting in wake contraction along the major axis. On average, the magnitude of the $\kappa_z$ component was 0.53 times that of $\kappa_y$ within the highlighted region, indicating comparatively weaker bending in the lateral direction. The spatial distribution of positive and negative $\kappa_z$ values was consistent with a spanwise contraction.

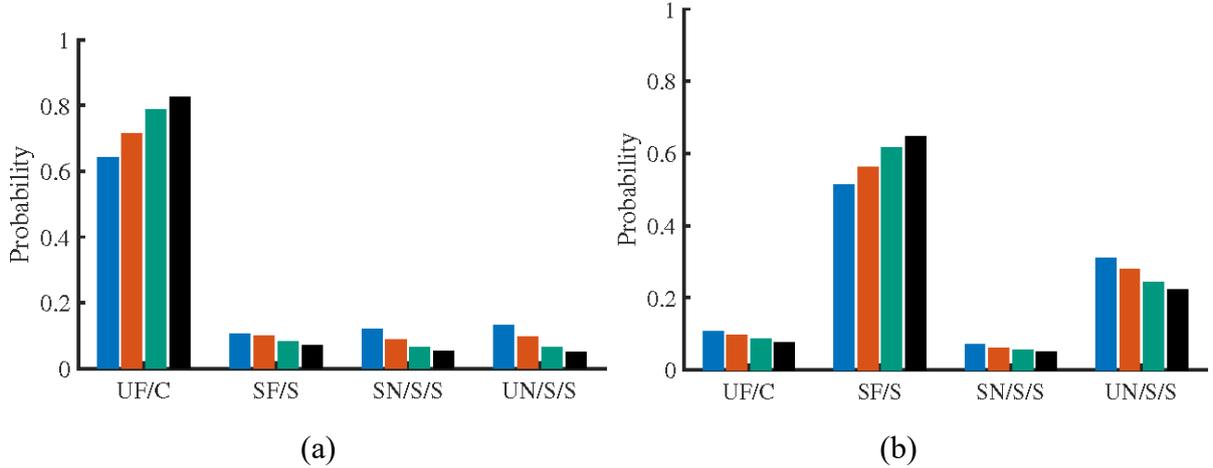

Figure 17: Probability distribution of the four incompressible flow topologies for the 5:1 ellipsoid, conditioned on (a) negative and (b) positive instantaneous enstrophy production. In (a) the colours correspond to regions where $\xi_{prod}/(U/D)^3 \leq -1$ (blue), $\leq -10$ (red), $\leq -50$ (green), and $\leq -100$ (black). In (b) the colours correspond to regions where $\xi_{prod}/(U/D)^3 \geq 1$ (blue), $\geq 10$ (red), $\geq 50$ (green), and $\geq 100$ (black).

Having established the central role of the intermediate eigenvalue and eigenvector in negative enstrophy production for the 5:1 ellipsoid, we now examine the local flow topology using data from an instantaneous snapshot. The relationship between the velocity invariants $Q$ and $R$ and local flow topology is described briefly in Appendix B. Figure 17 shows the prevalence of the four canonical topologies, namely, unstable focus / compressing (UF/C), stable focus / stretching (SF/S), stable node / saddle / saddle (SN/S/S), and unstable node / saddle / saddle (UN/S/S), conditioned on various thresholds of the instantaneous enstrophy production. Probability distributions for negative values of enstrophy production are shown in Figure 17(a), and those for positive production are shown in Figure 17(b). The results in Figure 17(a) indicate that negative enstrophy production is overwhelmingly dominated by the UF/C topology, with its prevalence increasing as production becomes more negative. The remaining three topologies contribute only marginally to negative production. Positive enstrophy production (Figure 17(b)) is dominated by the SF/S topology, along with a notable presence of the UN/S/S topology. As production becomes more positive, the prevalence of the SF/S topology increases while that of the UN/S/S topology decreases.

The correspondence between negative enstrophy production and local flow topology was examined by Buxton & Ganapathisubramani (2010) using stereoscopic particle image velocimetry of an axisymmetric turbulent jet. They found that enstrophy production was predominantly negative within flow regions characterized by the UF/C topology, and that regions associated with other flow topologies displayed a smaller presence of negative production, which is consistent with the observations in Figure 17(a). Bechlars & Sandberg (2017) also reported similar results from a turbulent boundary layer DNS, where persistent negative production was observed primarily in regions of UF/C topology. Overall, the results presented here indicate that the negative production region is dominated by the unstable focus / compressing topology, where the flow spirals outwards from a focus point and streams inwards along the axis of the spiral. This topological representation is consistent with earlier discussion of the role of the intermediate eigenvalue and eigenvector in negative enstrophy production; the physical interpretation of vorticity being aligned with the intermediate eigenvector, together with the intermediate eigenvalue being negative, is a vortex that is undergoing compression, which is consistent with the UF/C topology. We note that while the observations presented here hold for $Re_D = 10,000$, persistent negative enstrophy production was also observed near the poles of the 5:1 ellipsoid in a separate simulation at a lower $Re_D = 5,000$ (data not shown). Whether or not local negative production persists at significantly higher Reynolds numbers in bluff-body flows remains an open question.

## 4. Conclusion

In this work, we have conducted a simulation-based study of flow around prolate ellipsoids of various aspect ratios for a freestream Reynolds number of $Re_D = 10,000$ based on the minor axis diameter. The primary goal is to better understand the influence of body anisotropy on boundary layer separation, shear layer behavior, local flow topology, and enstrophy production in the wake. The ellipsoids were oriented at a 90° angle of attack with respect to the freestream, and the size and intensity of the recirculation bubble were observed to increase monotonically with increasing aspect ratio. The impact of body anisotropy on boundary layer detachment was examined using 2D cuts in the meridional and equatorial planes, as well as using 3D separation curves along the surface of the ellipsoids. The data indicate a delay in separation of the boundary layer in the meridional plane for larger *AR* values, but earlier separation in the equatorial plane. The decay rates of the streamwise velocity defect along the wake centerline were also examined, and the trends suggest a larger wake size, an extensive region of low wake pressure, and consequently, a higher drag coefficient for increasing body anisotropy.

Large body anisotropy gave rise to strong vortex tubes close to the equator, consistent with early breakdown of the shear layer in the equatorial plane. The strongest vortex tubes were observed at a distance approximately 2.5*D* downstream of the ellipsoid centers regardless of body anisotropy, which is related to positive enstrophy production reaching its maximum near this location. The resolved small-scale flow characteristics were examined in the wake using joint probability density functions of the second and third invariants of the velocity gradient tensor. A characteristic teardrop shape was observed consistently across all 5 ellipsoids, with minor influence of body anisotropy. The PDF data also indicated that vorticity dominates over strain in the wakes of all the ellipsoids, and that the vortex tubes are strongest for the highly anisotropic ellipsoid.

The 3D structures of positive and negative enstrophy production regions were visualized to better understand the connection of enstrophy production to local flow characteristics. Both positive and negative production exhibited shell-like structures, with contours of more intense production levels nested within contours of moderate production. The positive production contours separated out into two distinct lobes closer to the poles for higher body anisotropy. Regardless of body anisotropy, positive enstrophy production remained relatively low in the near-wake, and regions of highest positive production were observed in the intermediate wake approximately 2.5*D* downstream from the ellipsoids' centers. Negative enstrophy production was constrained to thin regions around the periphery of the highly anisotropic ellipsoids. Differently from positive production, negative production attained its maximum in the near-wake region. High body anisotropy in crossflow was shown to be more conducive to sustained negative enstrophy production in relatively large regions, owing to strong anisotropic contraction of streamlines in the cross-stream plane downstream of the high curvature poles. Enstrophy production remained positive beyond $x/D > 1.5$ for all ellipsoids independently of body anisotropy, indicating that the flow primarily undergoes vortex stretching in the intermediate- and far-wake. The contribution of the intermediate eigenvalue and eigenvector of the strain rate tensor was shown to be the primary source of intense negative production for the most anisotropic ellipsoid. The local flow topology was calculated to demonstrate that the negative production region is dominated by the unstable focus / compressing (UF/C) topology. This topological representation is consistent with the role of the intermediate eigenvalue and eigenvector in negative enstrophy production, where strong alignment of the vorticity vector with the intermediate eigenvector, together with a negative intermediate eigenvalue, leads to local vortex compression. We note that while the observations presented here hold for $Re_D = 10,000$, the existence of sustained negative production at significantly higher Reynolds numbers in bluff body flows requires further investigation.

# Acknowledgment

The authors gratefully acknowledge support from the Office of Naval Research (ONR) under contract number N00014-20-C-2035. This work was supported in part by high-performance computer time and resources from the DoD High Performance Computing Modernization Program.

# Declaration of Interests

The authors report no conflict of interest.

# Appendix

**Appendix A. Meshing strategy**

Computer Aided Design (CAD) models of the ellipsoids were imported into OpenFOAM to generate the corresponding computational mesh. The mesh resolution was kept sufficiently fine near the surface of the ellipsoid (wall) as well as in the near-wake region, to adequately capture large gradients in velocity. Additionally, the mesh was kept relatively coarse in regions of uniform flow to keep the computational cost manageable. Further refinement of the mesh in the volume slightly upstream of the ellipsoid and in the wake region of the body was used to correctly resolve small-scale flow structures that develop in these regions. An example cross-section of the computational mesh is shown in Appendix Figure 1. The total number of cells in the domain ranged from approximately 19 million to 32 million depending on the ellipsoid's aspect ratio. The minimum cell volume was $\approx 4.2 \times 10^{-8} D^3$, which corresponds to an equivalent cubic cell length of $3.5 \times 10^{-3} D$. The actual cell shapes were generalized hexahedrons and prisms. The finest cells were stretched close to the walls, to allow for improved resolution in the wall-normal direction.

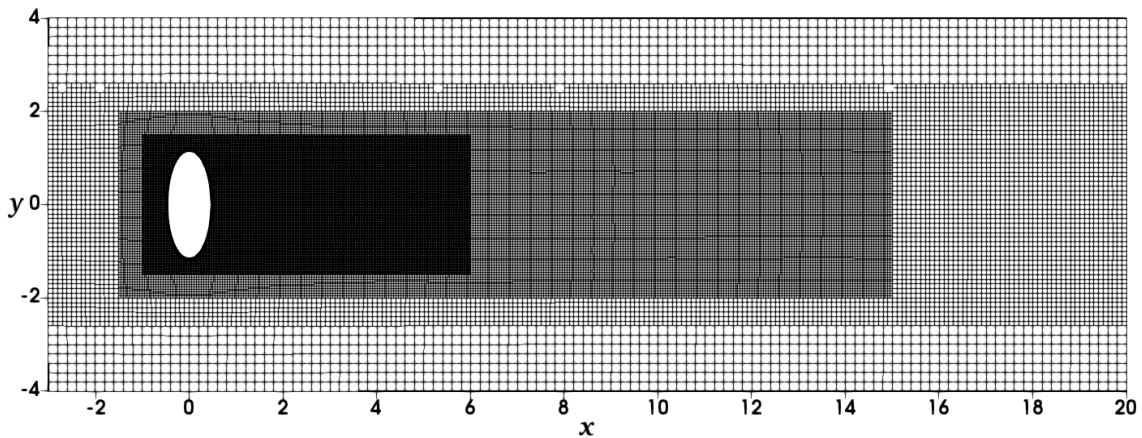

(a)

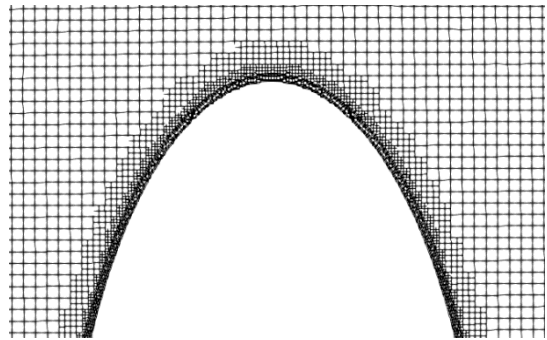

(b)

Appendix Figure 1: (a) View of the computational mesh in an *x-y* (meridional) plane cut through the domain, and (b) close-up view near the surface of the ellipsoid.

A mesh sensitivity analysis was performed using three different mesh configurations M1, M2 and M3, which are described in Appendix Table 1. The variation in results between M2 and M3 is minimal. However, conservatively, all simulations presented in this paper use mesh configurations that are better than the M3 mesh. The elevation separation angle ($\alpha$) corresponds to the intersection of the separation curve with the meridional plane. To ensure that separation curves correlate well with boundary layer detachment, the limiting streamlines and separation curve are visualized for the 5:1 ellipsoid in Appendix Figure 2.

**Appendix Table 1.** Mesh sensitivity analysis for the 5:4 ellipsoid.

|    | Total cells | Minimum cell volume | Drag coefficient ($C_D$) | Elevation separation angle ($\alpha$) |
|----|---|---|---|---|
| M1 | 3 million | $1.09 \times 10^{-7} D^3$ | 0.447 | 85.7 |
| M2 | 4.5 million | $6.55 \times 10^{-8} D^3$ | 0.442 | 86.1 |
| M3 | 7 million | $4.68 \times 10^{-8} D^3$ | 0.440 | 86.0 |

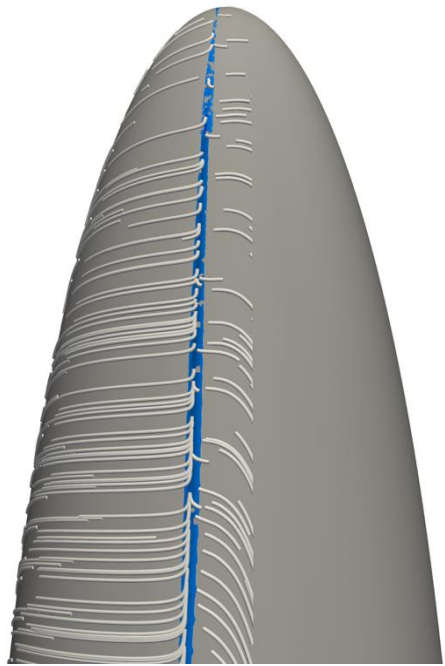

Appendix Figure 2: Limiting streamlines (white) and contour of zero wall shear stress magnitude (blue) for the 5:1 ellipsoid. The convergence of the streamlines signifies boundary layer separation, and it coincides with the zero-stress contour indicating that the contour (i.e., separation curve) provides a good indication of boundary layer separation.

The accuracy of the meshing strategy was assessed by plotting the first cell size in non-dimensional wall normal distance ($y^+$), and the fraction of resolved TKE ($k_{res}$) to the total TKE ($\gamma$ – Eq. (5)) in Appendix Figure 3. For flow around a bluff body in the subcritical regime, where the boundary layer remains laminar before undergoing separation, the flow in the wake region can be reasonably resolved with grid sizes of $y^+ \approx 5$. In the present computations, $y^+$ values on the ellipsoid surface were smaller than $y^+ \leq 2.5$ (Appendix Figure 3(a) and (b)). Due to separation of the flow, the value of $y^+$ decreases significantly at the downstream surface of the ellipsoid. The distribution of the resolved TKE fraction $\gamma$ for the flow around the ellipsoid is shown in Appendix Figure 3(c). It can be observed that more than 80% of TKE is resolved in the majority of the wake.

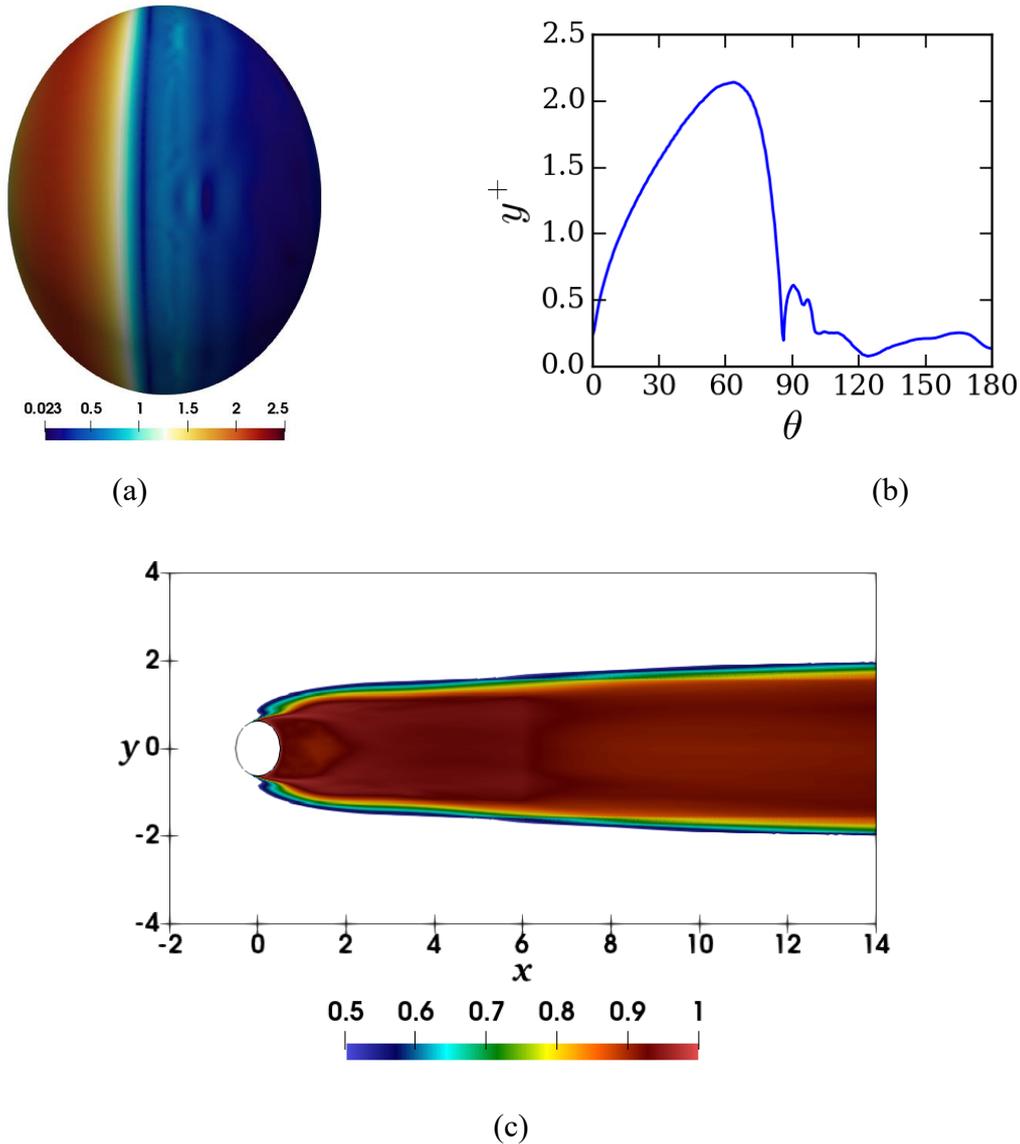

(a)

(b)

(c)

Appendix Figure 3: (a) Distribution of $y^+$ on the surface of the ellipsoid, and (b) variation of first cell-size $y^+$ along the azimuthal angle (here $\theta = 0°$ is at the upstream stagnation point). (c) Spatial distribution of the resolved TKE fraction ($\gamma$ – Eq. (5)).

To better understand the contribution of the subgrid model in various regions of the flow, the turbulent viscosity calculated via the dynamic model for the 5:1 ellipsoid is shown in Appendix Figure 4. The close-up image in Appendix Figure 4(a) indicates that the subgrid model has virtually no contribution in the near-wall layer, and minimal contribution in the near-wake region. Appendix Figure 4(b) indicates that the subgrid model's contribution increases modestly in the intermediate-wake due to high levels of positive enstrophy production (Figure 14), and to a greater extent in the far-wake where larger cell sizes were used to manage computational cost. From all of the observations presented in this section, it can be concluded that the present meshing and numerical strategies are able to predict the flow with reasonable accuracy.

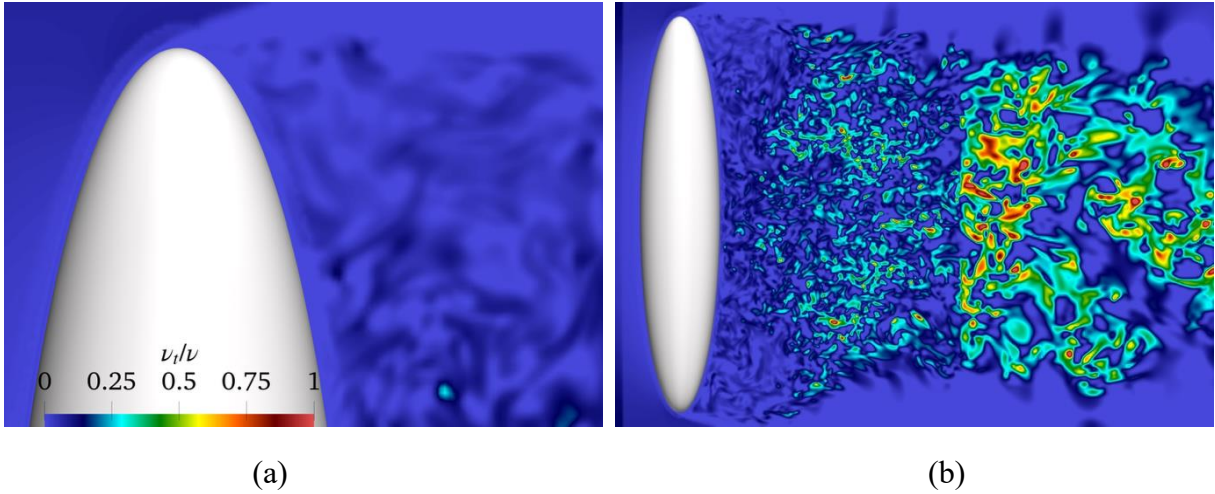

(a) (b)

Appendix Figure 4: (a) The ratio of turbulent viscosity to molecular viscosity is shown using a close-up view of the near-wake region, and (b) using a broader overview of the wake.

**Appendix B. Flow topology**

Each material fluid element in the flow can be considered to act as a critical point (i.e., a stagnation point) for an observer moving at the same velocity as the fluid element. The topological analysis is done in the context of this observer since the streamlines, and hence, the interpretation of local topology, can vary depending on an observer's reference frame velocity. The second invariant $Q = (\Omega_{ij}\Omega_{ij} - S_{ij}S_{ij})/2$ is the same as the $Q$ value used for visualization in Figure 11, where $\Omega_{ij}$ is the asymmetric part of the velocity gradient tensor and $S_{ij}$ is the symmetric part. The third invariant is $R = -\det(\nabla \boldsymbol{u})$, where $\boldsymbol{u}$ represents the velocity vector. Large positive values of $Q$ are associated with regions where vorticity dominates over strain, whereas negative values of $Q$ indicate that strain dominates vorticity. The sign of $R$ is related to the topology being classified as stable ($R<0$) or unstable ($R>0$). The discriminant $D_{QR} = Q^3 + \left(\frac{27}{4}\right)R^2$ is used to determine whether the eigenvalues of the velocity gradient tensor are real or complex. $D_{QR} < 0$ corresponds to all real eigenvalues and to strain-dominated regions, whereas $D_{QR} > 0$ corresponds to one real and a complex-conjugate pair of eigenvalues, and hence to rotation-dominated regions. Note that $Q>0$ implies necessarily that $D_{QR} > 0$, which indicates that regions where vorticity dominates over strain manifest as rotational tube-like structures, i.e., as vortex tubes. The dashed lines in Figure 12 indicate the curves where $R = 0$ and $D_{QR} = 0$, and they divide the phase space into four regions with distinct flow topologies. The visual and geometric interpretations of these flow topologies are discussed in detail by Soria et al. (1994) and by Ooi et al. (1999). The first quadrant with $R > 0, D_{QR} > 0$ corresponds to flow topology described as unstable focus / compressing (UF/C), where the flow tends to spiral outward from a critical point, while at the same time streaming inward along the longitudinal axis of the spiral to maintain the divergence-free condition. The second quadrant with $R < 0, D_{QR} > 0$ corresponds to stable focus / stretching (SF/S), where the flow spirals inward and streams outward along the longitudinal spiral axis. The third and fourth quadrants with $R < 0, D_{QR} < 0$ and $R > 0, D_{QR} < 0$, respectively, correspond to stable node / saddle / saddle (SN/S/S) and unstable node / saddle / saddle (UN/S/S) topologies. The node acts either as an attractor or a repeller depending on whether it is classified as stable or unstable. The saddle points attract flow along one axis and repel it along another orthogonal axis.

**Appendix C. Enstrophy transport equation**

The enstrophy transport equation is obtained by taking the dot product of vorticity with the vorticity form of the momentum equation:

$$\omega_i \left( \frac{D\omega_i}{Dt} = \omega_j \frac{\partial u_i}{\partial x_j} + \nu \frac{\partial^2 \omega_i}{\partial x_j \partial x_j} \right) \qquad \text{Eq. (8)}$$

$$\frac{D\xi}{Dt} = \omega_i S_{ij} \omega_j + \nu \frac{\partial^2 \xi}{\partial x_j \partial x_j} - \nu \frac{\partial \omega_i}{\partial x_j} \frac{\partial \omega_i}{\partial x_j} \qquad \text{Eq. (9)}$$

The inviscid enstrophy production term $\omega_i S_{ij} \omega_j$ depends on the interaction of the vorticity vector with the strain-rate tensor, and locally it can be either negative or positive. Globally, net enstrophy production is positive for turbulent flows due to the predominance of vortex stretching in the forward cascade process. Positive production leads to enstrophy generation via vortex stretching, but production can also be negative locally where vortex compression leads to enstrophy depletion (Buxton & Ganapathisubramani, 2010). This local reduction in enstrophy does not reduce the net enstrophy in the domain, and it may be related to a potential backscatter mechanism in the energy cascade (Bechlars & Sandberg, 2017). We distinguish between enstrophy 'depletion' and enstrophy 'dissipation' here since negative production does not cause a global dissipative reduction of total enstrophy in the domain. The viscous enstrophy dissipation term $\nu \frac{\partial \omega_i}{\partial x_j} \frac{\partial \omega_i}{\partial x_j}$ leads to net enstrophy destruction in the domain, with motion being dissipated in the form of heat due to viscous effects. The other viscous term, i.e., the enstrophy diffusion term $\nu \frac{\partial^2 \xi}{\partial x_j \partial x_j}$ reduces enstrophy gradients by redistributing enstrophy spatially in the domain. Although it is viscous in nature, the diffusion term can be represented in conservation form (i.e., as a divergence term), and it does not act as a global sink for total enstrophy in the domain.